\documentclass[reprint, amsmath, amssymb, aps,prd,superscriptaddress]{revtex4-2}
\usepackage{graphicx} 
\usepackage{physics}
\usepackage{xcolor}
\usepackage[colorlinks=false,pdfborder={0 0 0}]{hyperref}
\usepackage{soul}
\usepackage{float}
\usepackage{ragged2e}
\usepackage[caption=false]{subfig}

\newcommand{\manyfiguresb}[9]{
    \begin{figure*}[t]
        \centering

        \subfloat[]{
            \begin{minipage}[c]{0.47\textwidth}
                \centering
                \includegraphics[width=\textwidth]{#1}
            \end{minipage}
        }
        \hfill
        \subfloat[]{
            \begin{minipage}[c]{0.47\textwidth}
                \centering
                \includegraphics[width=\textwidth]{#2}
            \end{minipage}
        }

        \subfloat[]{
            \begin{minipage}[c]{0.47\textwidth}
                \centering
                \includegraphics[trim={#8},clip,width=#9]{#3}
            \end{minipage}
        }
        \hfill
        \subfloat[]{
            \begin{minipage}[c]{0.47\textwidth}
                \centering

                \includegraphics[width=0.345\textwidth]{#4}

                (d1)

                \vspace{0.3cm}

                \includegraphics[width=0.345\textwidth]{#5}

                (d2)

            \end{minipage}
        }

        \caption{#6}
        \label{#7}
    \end{figure*}
}

\begin{document}

\preprint{APS/123-QED}

\title{Unruh-DeWitt detectors in a delocalized massive shell}

\author{Giovanna F. do Valle}
\thanks{These authors contributed equally to this work.}
\affiliation{Departamento de F\'isica - ICEx, Universidade Federal de Minas Gerais, CP 702, 30161-970, Belo Horizonte, MG, Brazil}

\author{Saadat Salman Shariff}
\thanks{These authors contributed equally to this work.}
\affiliation{Research Center for Quantum Information, Institute of Physics, Slovak Academy of Sciences, D\'ubravsk\'a Cesta 9, 84511 Bratislava, Slovakia}

\author{Natália S. Móller}
\email{Contact author: natalia.smoller@uniroma2.it}
\affiliation{Research Center for Quantum Information, Institute of Physics, Slovak Academy of Sciences, D\'ubravsk\'a Cesta 9, 84511 Bratislava, Slovakia}
\affiliation{Dipartimento di Fisica, Universit\`a di Roma Tor Vergata, Via della Ricerca Scientifica 1, 00133, Roma, Italy}

\author{Nelson Yokomizo}
\email{Contact author: yokomizo@fisica.ufmg.br}
\affiliation{Departamento de F\'isica - ICEx, Universidade Federal de Minas Gerais, CP 702, 30161-970, Belo Horizonte, MG, Brazil}

\date{\today}

\begin{abstract}
We construct a model for the response of a detector to vacuum fluctuations inside a massive hollow shell in a superposition of positions. The delocalized source mass gives rise to a superposition of spacetime geometries and their associated vacuum fluctuations. To describe this system, we employ the formalism of quantum reference frames to transition to a frame in which the source mass occupies a definite position while the detector is in a superposition of positions. This allows the dynamics of the detector to be analyzed using standard tools of quantum field theory on a definite curved spacetime. We compute the detector response for a wide range of configurations of the shell and detector positions and investigate how information about the quantum state of the shell is encoded in the response function, identifying features of the superposition that are effectively captured by the detector. In particular, we show that, under appropriate postselection, the response function exhibits interference patterns, and identify regions of the parameter space of the model in which the shell-induced contributions to the interference terms are maximized.

\bigskip

\end{abstract}

\maketitle

\section{Introduction}

In the absence of a full theory of quantum gravity, phenomenological models based on simple assumptions about how quantum effects manifest themselves in gravitational settings have been developed to investigate possible signatures of a nonclassical behavior of the geometry of spacetime, especially in the regime of weak gravitational fields. In a typical scenario, one studies the influence of a nonclassical gravitational field produced by a mass in a superposition of positions on the evolution of another quantum system, which serves as a probe of the nonclassical geometry. Predicted effects at the interface of quantum mechanics and gravity include a universal mechanism for the decoherence of extended quantum systems~\cite{Zych:2011hu,Pikovski:2013qwa}, gravitationally-induced entanglement production~\cite{Bose:2017nin,Marletto:2017kzi}, appearance of repulsive gravitational forces under postselection~\cite{Pablo2026}, and processes with indefinite causal order~\cite{Zych:2017tau,Moller:2023iwo}. 

Low-energy quantum systems in a superposition of geometries have been considered in several works. In this regime, one can design simple setups in which a source mass prepared in a delocalized state interacts gravitationally with a probe quantum system prepared in a localized state or in a simple superposition of localized states; for reviews of the area, see \cite{Sahdo_Moller_2025,RevModPhys.97.015003,Marletto:2024ltk}. The probe system then evolves under a superposition of proper time flows, and its internal state becomes entangled with that of the source mass.  The superposition of proper times can be revealed through suitable measurements of the final state of the probe or through the resulting decoherence, and constitutes a key mechanism underlying the emergence of low-energy quantum gravity effects \cite{Christodoulou:2018cmk}.

The case of quantum fields is less studied and poses new conceptual and technical questions. Extending the approach employed for low-energy systems to the description of quantum fields in a superposition of geometries first requires a proper definition of the states and observables in this regime. Regarding the definition of the states, one immediately encounters the difficulty that these are inherently global objects and, in general, cannot be interpreted as being spatially localized \cite{Haag:1996hvx}. In particular, the definition of the vacuum state depends on the global structure of spacetime, and it is not obvious how it should be generalized for nonclassical spacetimes, even in the relatively simple settings typically considered in quantum gravity phenomenology. In addition, the mathematical description of local field operators must also be extended to account for field measurements on a superposition of geometries. 

Two main approaches have been explored in recent works to model the behavior of quantum fields on simple superpositions of geometries, with applications to scalar fields without self-interactions. In a series of works \cite{FooPRD2023,GoelPRD2025,FooPRL2022}, superpositions of quotients of a common classical spacetime, obtained through distinct identifications of its points, were considered. States and field operators on the superposition of geometries were defined from combinations of the corresponding quantities on the original spacetime, allowing the properties of the quantum field on the superposition of geometries to be computed by working in the original classical spacetime. This strategy has been applied to superpositions of cylindrically identified Minkowski spacetimes \cite{FooPRD2023, GoelPRD2025} and BTZ black holes with different masses \cite{FooPRL2022}. An alternative strategy consists of applying a change of quantum reference frame (QRF) \cite{angelo2011physics,Giacomini:2018gmq} that maps a model involving a quantum field on a superposition of geometries to an equivalent model on a classical spacetime, and has been applied to a superposition of translated de Sitter spacetimes \cite{Foo_2021} and BTZ black holes \cite{CepollaroBTZ2026}. In both approaches, quantum fluctuations of the field are probed through the response of a particle detector coupled to the field, modelled as an Unruh--DeWitt (UDW) detector \cite{Unruh:1976db,DeWitt:1980hx}.

In this work, we analyze a new setup involving a scalar field on the nonclassical gravitational field of a spherical shell in a superposition of distinct locations. The separation between the possible locations of the shell is small enough that there exists a common interior region lying inside the shell in either location. An Unruh-DeWitt detector is placed in this region. Under a change of quantum reference frame, the model is mapped into an equivalent one in which the shell is at a definite location while the detector is delocalized within its interior. Such an equivalent setup can be analyzed within the framework of delocalized detectors introduced in \cite{FooPRD2020} and applied in several previous works \cite{FooPRD2020,PhysRevD.103.065013,PhysRevD.102.045002,Paczos:2023wsy}. The normal modes of a scalar field on the geometry of a spherical shell can be determined analytically and have a sufficiently simple form to enable the numerical evaluation of the detector response for a wide range of spatial configurations.

In contrast to previous analyses of quantum fields on superpositions of spherically symmetric geometries \cite{FooPRL2022,CepollaroBTZ2026}, which considered BTZ black holes in three spacetime dimensions, our analysis is carried out directly in four spacetime dimensions, without resorting to a lower-dimensional simplification. By considering a superposition of spherical shells rather than black hole geometries, we obtain a setup that is simple enough to allow the numerical evaluation of the detector response in the physical dimensionality of spacetime. Moreover, when the relative separation between the shell positions is small, the resulting geometry contains an interior region that is identical in both branches---namely, a local patch of flat spacetime---so that the branches cannot be distinguished by classical measurements within this region.

The case of a localized UDW detector inside a spherical shell at a definite location was previously studied in \cite{Ng:2016hzn}. It was shown that, although the detector lies in a locally flat region of spacetime, its response can distinguish between globally flat Minkowski spacetime and a spacetime containing a massive spherical shell, even when the detector is switched on for a time interval shorter than that required for a signal to travel from the detector to the shell and back. This shows that the detector can be influenced by the presence of the shell even when no curvature can be detected by classical means. Similarly, in our setup the detector lies in a locally flat region common to both branches of the superposition. We consider time scales for which the shell cannot be detected in either branch, as well as more general configurations. This allows us to address the more general question of whether the detector can distinguish, solely from the influence of the delocalized shell on the state of the field, between a spacetime generated by a shell at a definite position and a nonclassical spacetime produced by the same shell in a superposition of positions.

This work is organized as follows. In Section II, we review the formalism of quantum-controlled Unruh-DeWitt detectors, applied both for a detector in a superposition of positions and for a nonclassical geometry produced by a mass in superposition of positions. These two setups are formulated so that they are related by a change of quantum reference frame. In Section III, we apply this formalism to derive analytical formulas for the response of a Unruh-DeWitt detector in the geometry produced by a massive thin shell in a superposition of two positions. The response function is then numerically computed in Section IV for all possible configurations such that the detector lies in the flat region inside the shell for both branches of the superposition. The results are summarized and discussed in the Section V.

\section{Quantum-controlled UdW detectors}
In this section, we first review the standard formulation of the UDW particle detector model, an operational way of modelling the interaction between a two-level system and a quantum field. We then review the description of a UDW detector in a superposition of paths and introduce a model for a UDW detector interacting with a gravitational field produced by a massive object in a superposition of locations. The latter is formulated so that, under a change of quantum reference frame, it is equivalent to a model in which the massive object is at a definite location while the detector is in a superposition of paths.

\subsection{UDW detector}
The UDW detector is perhaps the simplest model of a local detector, providing an operational definition of the particle content of a quantum field. It was first introduced by Unruh \cite{Unruh:1976db} and later simplified by DeWitt \cite{DeWitt:1980hx}. It is a two-level, point-like quantum mechanical system with an energy gap $\Omega$ moving through spacetime along a worldline $\textbf{x}(\tau)$ parametrised by its proper time $\tau$. While traversing this trajectory, it interacts with a local scalar field $\hat{\phi}(x)$ and can get excited or decay due to the interaction with the field. We consider a free massless real scalar field $\hat{\phi}(x)$.

Let $\ket{0}$ and $\ket{1}$ be the ground and excited states of the detector, respectively. Its free Hamiltonian is
\begin{equation}\label{eq:def_free_detector_hamiltonian}
\hat{H}_0=\frac{\Omega}{2}\left(\ket{1}\bra{1}-\ket{0}{}\bra{0}\right).
\end{equation}
The interaction Hamiltonian is given by
\begin{equation}\label{eq:interaction_hamiltonian}
\hat{H}_I(\tau)=\lambda\chi(\tau)\hat{\mu}(\tau)\hat{\phi}(x(\tau)),
\end{equation}
where $\lambda\ll1$ is the dimensionless coupling constant that determines the strength of the interaction between the detector and the quantum field, $\chi(\tau)$ is the switching function that determines when and for how long the detector interacts with the field, and $\hat{\mu}(\tau)$ is the monopole moment operator governing the transition between the internal states of the detector. In the interaction picture, this operator is given by
\begin{equation}
\hat{\mu}(\tau)=\mathrm{e}^{\mathrm{i} \hat{H}_{0}\tau} \hat{\mu}(0) \mathrm{e}^{-\mathrm{i} \hat{H}_{0} \tau}=\mathrm{e}^{\mathrm{i} \Omega \tau} \hat{\sigma}^{+}+\mathrm{e}^{-\mathrm{i} \Omega \tau} \hat{\sigma}^{-},
\end{equation}
where $\hat{\sigma}^{+} =|1\rangle\langle 0|$ and $\hat{\sigma}^{-}=|0\rangle\langle1|$.

The composite system formed by the detector and the scalar field is prepared in the initial state $\hat{\rho}_{i}=\ket{0}{}_D\bra{0}\otimes\ket{0}{}_\phi\bra{0}$ before the interaction is turned on. The time evolution of the total system is described by the unitary operator
\begin{equation}\label{eq:def_unitary_operator}
\hat{U}=\hat{\mathcal{T}} \exp \left(-i \int_{-\infty}^{\infty} \mathrm{d} \tau \hat{H}_I(\tau)\right).
\end{equation}
Once the interaction is turned off, the final state of the detector-field system is $\hat{\rho}_{f}=\hat{U}\hat{\rho}_i\hat{U}^{\dagger}$. By tracing out the field degrees of freedom, we obtain the final state of the detector to the leading order:
\begin{equation}
    \hat{\rho}_{D}=\mathrm{Tr}_\phi[\hat{U}\hat{\rho}_i\hat{U}^{\dagger}] = \left(\begin{array}{cc}
1-P_D & 0 \\
0 & P_D
\end{array}\right)+\mathcal{O}\left(\lambda^4\right)
\end{equation}
in the ordered basis $\{\ket{0}_D,\ket{1}_D\}$. Here, $P_D$ is the transition probability of the detector to be found in the excited state $\ket{1}_D$. It can be expressed as follows:
\begin{equation}
P_D=\lambda^2|_D\langle 0| \hat{\mu}(0)|1\rangle_D|^2 \mathcal{F}(\Omega) \, ,
\end{equation}
where the factor $\lambda^2|{ }_D\langle 0| \hat{\mu}(0)|1\rangle_D|^2$ depends only on the properties of the detector, whereas the response function
\begin{align}
\mathcal{F}(\Omega)=&\int_{-\infty}^{\infty} d \tau_1 \int_{-\infty}^{\infty} d \tau_2~ \chi(\tau_1) \chi(\tau_{2}) e^{-i \Omega\left(\tau_2-\tau_1\right)}\\\nonumber &\times\mathcal{W}\left(x(\tau_2), x\left(\tau_1\right)\right)
\end{align}
also depends on the structure of spacetime through the Wightman function (two-point correlation function) evaluated along the worldline of the detector,
\begin{equation}
    \mathcal{W}\left(x(\tau_2), x\left(\tau_1\right)\right)={}_\phi\bra{0}\phi(x(\tau_2))\phi(x(\tau_1))\ket{0}_\phi \, .
\end{equation}

\subsection{UDW detector in a superposition of paths}\label{sec:UDW_control}

We now generalize the UDW model to the case in which the detector is in a superposition of different positions. This can be done by introducing a control degree of freedom $C$ to represent the position of the detector~\cite{Foo:2020dzt}. The Hilbert space of the combined system is
\begin{equation}
\mathcal{H}=\mathcal{H}_{D}\otimes\mathcal{H}_{C}\otimes\mathcal{H}_{\phi} \, ,
\end{equation}
where $\mathcal{H}_{D}$, $\mathcal{H}_{C}$ and $\mathcal{H}_{\phi}$ describe the internal degrees of freedom of the detector, its position, and the state of the field, respectively. We are interested in the case in which the detector is prepared in a superposition of two positions. The control system is then described by the coherent superposition
\begin{equation}
 \ket{X}_C=\frac{\ket{X_1}+\ket{X_2}}{\sqrt{2}} \, ,
\end{equation}
where the orthogonal states $\ket{X_1}$ and $\ket{X_2}$ correspond to the two different locations of the detector. 

Let the initial state of the full system be
\begin{equation}
\hat{\rho}_{i,D}=\ket{0}_{D}\bra{0}\otimes\ket{0}{}_{\phi}\bra{0}\otimes\ket{X}{}_{C}\bra{X} \, .
\label{eq:initial-state-D}
\end{equation}
Upon introducing the quantum control, the interaction Hamiltonian in Eq.~\eqref{eq:interaction_hamiltonian} becomes
\begin{equation}
    \hat{H}_\text{I}^{(D)}(\tau) = \sum_{i=1}^{2}\hat{H}_i^{(D)}(\tau)\otimes \ket{X_i}{}_{C}\bra{X_i},
\label{eq:interaction_hamiltonian_superposed}
\end{equation}
where
\begin{equation}
\hat{H}_i^{(D)}(\tau)=\lambda\chi(\tau)\hat{\mu}(\tau)\hat{\phi}(x_i(\tau))
\end{equation}
governs the interaction of the detector and the scalar field along the worldline $x_i(\tau)$ associated with the $i$-th branch of the superposition. The time evolution of the combined control-detector-field system now becomes conditioned on the position of the detector and takes the form
\begin{equation}
    \hat{U}^{(D)}=\sum_{i=1}^{2}\hat{U}_i^{(D)}\otimes \ket{X_i}{}_{C}\bra{X_i},
\end{equation}
where
\begin{equation}\label{eq:def_unitary_operator_branch}
\hat{U}_i^{(D)}=\hat{\mathcal{T}} \exp \left(-i \int_{-\infty}^{\infty} \mathrm{d} \tau \hat{H}_i^{(D)}(\tau)\right).
\end{equation}
The perturbative expansion of the evolution operator for the $i$-th branch is
\begin{align}
    \hat{U}_i^{(D)}= \mathbb{I}&+(\mathrm{-i})\int \mathrm{d} \tau_1 \hat{H}_i^{(D)}(\tau_1) \\\nonumber&+\frac{(\mathrm{-i})^2}{2} \int \mathrm{d} \tau_2  \mathrm{d} \tau_1 \hat{H}_i^{(D)}(\tau_2) \hat{H}_i^{(D)}\left(\tau_1\right)
    +\mathcal{O}\left(\lambda^3\right).
\end{align}

Suppose that in the far future the control is measured in the diagonal basis
\begin{equation}
    \ket{X_{\pm}}=\frac{\ket{X_1}\pm\ket{X_2}}{\sqrt{2}} \, .
\end{equation}
The reduced density matrix of the detector, after tracing out the field, is then
\begin{equation}
    \hat{\rho}_D = \begin{pmatrix} 1 - P_E^{(\pm)} & 0 \\ 0 & P_E^{(\pm)} \end{pmatrix},
\end{equation}
where the transition probability is given by 
\begin{equation}
\label{eq:transition-pe}
P_E^{(\pm)}=\frac{\lambda^2}{4} \left[\mathcal{F}_{11}(\Omega)+\mathcal{F}_{22}(\Omega)\pm2\mathcal{F}_{12}(\Omega)\right],
\end{equation}
with
\begin{align}\label{eq:response_fn_ij_def}
\mathcal{F}_{ij}^{(D)}(\Omega)=&\int_{-\infty}^{\infty} d \tau_1 \int_{-\infty}^{\infty} d \tau_2 \chi\left(\tau_1\right) \chi\left(\tau_2\right) e^{-i \Omega\left(\tau_2-\tau_1\right)} \nonumber\\&\times\mathcal{W}^{(D)}_{ij}\left(x_i(\tau_2), x_j\left(\tau_1\right)\right) \, ,
\end{align}
and 
\begin{equation}\label{eq:wightman_fn_ij_def}
    \mathcal{W}^{(D)}_{ij}\left(x_i(\tau_2), x_j\left(\tau_1\right)\right)=\langle 0| \hat{\phi}\left(x_i\left(\tau_2\right)\right) \hat{\phi}\left(x_j\left(\tau_1\right)\right)|0\rangle \, .
\end{equation}
Here, $\mathcal{F}_{11}^{(D)}(\Omega)$ and $\mathcal{F}_{22}^{(D)}(\Omega)$ are the local contributions to the transition probability coming from the same branch of the superposition. Unlike a detector traveling along a single classical worldline, $P_E^{(\pm)}$ now has non-local contributions $\mathcal{F}_{12}^{(D)}(\Omega)$ coming from the cross-branch Wightman function.

\subsection{UDW in a superposition of geometries}

In this section, we extend the UDW model to a scenario where the detector interacts with a gravitational source in a spatial superposition. Specifically, we consider a massive object prepared in a superposition of two distinct locations, without initially specifying its particular geometry. In our setup, this superposition is modeled as a rigid spatial translation, ensuring that the intrinsic geometric properties of the mass are the same in both branches. We employ a common coordinate system for the two branches and consider a situation in which a specific subregion has the same local geometry in both branches. The detector is placed within this region, so that the local metric, and consequently the notions of spatial distance and proper time in its neighborhood, are the same in both branches and independent of the mass’s position.

The evolution of the internal state of the UDW detector in a nonclassical geometry produced by a mass in a superposition of positions is described using a control degree of freedom, analogously to the framework introduced in Sec.~\ref{sec:UDW_control}. For a mass prepared in a superposition of two positions, the control degree of freedom is associated with two orthogonal states, $\ket{M_1}$ and $\ket{M_2}$, corresponding to the two possible locations of the mass. It is important to note that a rigorous definition of a Hilbert space for generic superpositions of spacetimes remains an open question. However, this does not preclude the analysis of the proposed scenario, which we regard as a phenomenological model to explore consequences of a superposition of geometries for the interaction of a quantum field with a localized detector.

The properties of the field at a specific point depend on the position of the mass. We consider an initial state of the mass and the field given by
\begin{equation}
    \ket{\psi_M} = \frac{1}{\sqrt{2}}\sum_{i=1}^2 \ket{M_i}\otimes \ket{0_i} \, ,
\end{equation} 
where $\ket{0_i}$ denotes the vacuum state of the field on the $i$-th geometry. In such a state, the local vacuum fluctuations of the field at a given position are controlled by the state of the mass. Considering the detector prepared in its ground state, the initial state of the full system composed of the detector, the mass, and the field is then
\begin{equation}
    \hat\rho_{i, M} = \ket{0}_D\bra{0}\otimes \ket{\psi_M}\bra{\psi_M} \, .
\label{eq:initial-state-M}
\end{equation}

We assume that the interaction of the detector with the field is captured by the interaction Hamiltonian 
\begin{equation}
    \hat H_I^{(M)}(\tau) = \sum^2_{i=1} \hat{H}_i^{(M)}(\tau)\otimes\ket{M_i}_C\bra{M_i},
\label{eq: Hamiltonian_mass}
\end{equation}
where
\begin{equation}
    \hat H_i^{(M)}(\tau) = \lambda \chi(\tau)\hat\mu(\tau)\hat\phi_i(x(\tau)) \, .
\end{equation}
In this model, in each branch of the superposition of geometries, the detector responds to the field as in the corresponding classical geometry. The label $i$ in the field operator $\hat{\phi}_i(x)$ means the field operator at the position $x$ in the representation associated with the $i$-th geometry. Note that, unlike in Eq.~\eqref{eq:interaction_hamiltonian_superposed}, the field operator $\hat\phi_i$ is evaluated at a common coordinate position $x$ for both configurations associated with the mass's locations. However, the position of the shell varies between the states in superposition, which results in the vacuum fluctuations at the location of the detector being dependent on the position of the mass. At this point the construction must be understood as a formal procedure, since we have not yet specified how the field operators $\hat{\phi}_i$ act on states on a superposition of spacetimes. We will soon define the relevant cross-branch matrix elements of these operators through a change of quantum reference frame.

The calculation of the transition probability is similar to those performed in the frameworks considered in the previous sections. After measuring the control in the basis $\left(\ket{M_1}\pm\ket{M_2}\right)/\sqrt{2}$ and tracing over the field degrees of freedom, the transition probability has again the form described in Eq.~\eqref{eq:transition-pe}, but now with contributions $\mathcal{F}^{(M)}_{ij}$ given by
\begin{multline}
    \mathcal{F}_{ij}^{(M)}(\Omega) = \int^\infty_{-\infty} d\tau_1 \int^\infty_{-\infty} d\tau_2 \chi(\tau_1)\chi(\tau_2) e^{-i\Omega(\tau_2-\tau_1)}\\
    \times \mathcal{W}_{ij}^{(M)}(x(\tau_2),x(\tau_1))
    \label{eq:fijM}
\end{multline}
where 
\begin{equation}
\label{eq:wightmann-cross-branch}
    \mathcal{W}_{ij}^{(M)}(x(\tau_1),x(\tau_2)) = \bra{0_i}\hat\phi_i(x(\tau_2))\hat\phi_j(x(\tau_1))\ket{0_j}.
\end{equation}

A key distinction from the case of a superposition of paths is that the contribution $\mathcal{F}_{12}^{(M)}$ now involves a correlation between different vacuum states, $\ket{0_1}$ and $\ket{0_2}$, each associated with a distinct geometry. Evaluating the cross-branch correlation function $\mathcal{W}_{12}^{(M)}$ poses a significant challenge, since  standard quantum field theory in curved spacetimes does not provide a prescription for defining correlations between field operators and states on superposed spacetimes. 

To overcome this lack of formal machinery, we exploit the specific symmetries of our setup. Since the two branches of the superposition are related by a rigid spatial translation, we can employ a simple transformation of quantum reference frames. As we will show in the next section, this transformation allows us to map the scenario of a mass in superposition to an equivalent one in which the detector is in a superposition of trajectories on a single, definite geometry. In this description, there is a single vacuum state, and the two-point functions become ordinary, well-defined Wightman functions, providing a straightforward way to evaluate the detector's response.

\subsection{QRF transformation}

In standard quantum field theory, reference frames are typically treated as classical, idealized systems external to the quantum system under consideration. However, a reference frame is, in general, a physical system. For example, a collection of rods can be used to construct a system of spatial coordinates. The formalism of quantum reference frames (QRFs) was introduced to describe reference frames associated with quantum rather than classical systems~\cite{angelo2011physics, Eddington1949,Aharonov1967,Aharonov1984,Bartlett2007}. It extends the classical notion of a change of reference frame to transformations between quantum reference frames, which can involve superpositions of classical changes of coordinates.
For example, suppose that a composite system, say a massive body and a particle, is prepared in a configuration where the mass is in a delocalized state and the particle is in a localized state. Under a suitable QRF transformation, one can instead describe the mass in a localized state and the particle in a delocalized state.

The theoretical landscape of QRFs includes several distinct formalisms~\cite{Giacomini:2018gmq, vanrietvelde2020change, Loveridge2018, extraparticle}. Although the connections among these formalisms are still under investigation~\cite{castroruiz2025interpreting}, different QRF frameworks have already been applied to a variety of problems in both gravitational and nongravitational settings
\cite{LeonTime, Hoehn2020, Castro-Ruiz:2019vsu, Trinity, FlaminiaEinsteinEqPrinc,  Vik2023, ACVik2023, Bojowald2023, MichaelPaiva2024, Cepollaro2024, devuyst2025relation, Ladina2025, Mattei2026, Chataignier2026, CepollaroBTZ2026, Bruna2026, Miquel2025, QuantumGroupsQRF}.
Of particular interest to our purposes, changes of quantum reference frames have been employed to formulate the dynamics of quantum systems in spacetimes with indefinite metric produced by a mass in a superposition of positions \cite{ACVik2023, CepollaroBTZ2026, Castro-Ruiz:2019vsu}. The general strategy consists of first changing to a frame in which the source mass is at a definite position. In this frame, the spacetime geometry is definite, and the dynamics of the quantum system of interest can be studied using standard techniques. Assuming that the physical laws are invariant under changes of quantum reference frames, the resulting description can then be translated to the original frame, in which the metric is indefinite. This allows one to sidestep the fundamental problem of constructing a general description of the space of quantum states of gravity and investigate specific setups at the interface of quantum mechanics and gravity.

The perspectival framework \cite{Giacomini:2018gmq} is particularly well-suited to implementing this strategy for a mass in a superposition of configurations related by rigid spatial translations, as done in  \cite{ACVik2023, CepollaroBTZ2026, Castro-Ruiz:2019vsu}. We adopt the same approach to provide a concrete formulation of the dynamics of a UDW detector in the superposition of geometries introduced in the previous section. Since our setup involves translated geometries with a common isometric subregion around the detector, the notions of spatial distance and proper time in the vicinity of the detector remain consistent across the branches in superposition. This allows us to identify corresponding neighborhoods of the detector in the different frames.

Let us now describe how the models for a UDW detector in a superposition of geometries and in a superposition of paths are related by an explicit change of quantum reference frame in the perspectival approach. We begin in the frame in which the mass sourcing the gravitational field is in a superposition state, corresponding to the frame with an indefinite metric. Let $\hat T_i$ denote the translation that maps the position $\mathbf{R}_i$ of the mass in the configuration $\ket{M_i}$ to the origin of the coordinates, and let $\mathbf{x}$ denote the position of the detector in the original frame. The transformation between the two quantum reference frames is mediated by the operator
\begin{equation}\label{eq:qrf_transformation_def} 
    \hat S = \sum^2_{i=1} \ket{X_i}\bra{M_i}\otimes \hat T_i \otimes \mathbf{1}_D,
\end{equation}
where $\ket{X_i}$ denotes the state of the control system representing the detector at the position $\mathbf{x}_i=\mathbf{x} - \mathbf{R}_i$ in the new reference frame, in which the shell is centered at the origin of the coordinates. The factor $\mathbf{1}_D$ ensures that the internal state of the detector is unaffected by the transformation. The action of the translation $\hat T_i$ on the field states is such that
\begin{equation}
\ket{0} = \hat T_i \ket{0_i} \, ,
\end{equation}
where $\ket{0}$ is the vacuum state when the shell is centered at the origin of the coordinates, while $\ket{0_i}$ is the vacuum state when the mass shell is centered at $\mathbf{R}_i$, corresponding to the $i$-th branch of the superposition of geometries. The operator $\hat T_i$ shifts the superposition dependency from the field configuration to the coordinate argument, such that
\begin{equation}
\hat T_i \hat \phi_i(\mathbf{x})\hat T_i^\dagger = \hat \phi(\mathbf{x}_i) \, .
\label{eq:field-transformation}
\end{equation}

Applying the transformation to the initial state $\hat\rho_{i, M}$ in which the mass is in superposition yields:
\begin{align}    \label{eq:change-q-frame}
\hat{S}\hat{\rho}_{i, M}\hat{S}^\dagger 
    &= \ket{0}_D\bra{0} \otimes \ket{0}_\phi\bra{0} \otimes \ket{X}_C\bra{X}
    \nonumber
\\
    & = \hat{\rho}_{i, D} \, ,
\end{align}
where $\ket{X}_C = (\ket{X_1} + \ket{X_2})/\sqrt{2}$. That is, we obtain the initial state \eqref{eq:initial-state-D} introduced in Sec.~\ref{sec:UDW_control}, in which the detector is in a superposition of trajectories. 

We establish the operational equivalence of the dynamics in the two models by transforming the Hamiltonian \eqref{eq: Hamiltonian_mass} for the scenario with a superposition of masses under the change of quantum reference frame defined by Eq.~\eqref{eq:qrf_transformation_def} and verifying that the resulting Hamiltonian coincides with the Hamiltonian \eqref{eq:interaction_hamiltonian_superposed} for the scenario with a superposition of paths. The transformation yields
\begin{equation}
    \hat S \hat H^{(M)}_I\hat S^\dagger = \sum ^2_{i=1} \left(\hat T_i \hat H_{i}^{(M)}\hat T^\dagger_i\right) \otimes  \ket{X_i}_C\bra{X_i} \, .
\end{equation}
It follows from Eq.~\eqref{eq:field-transformation} that
\begin{align}
    \hat S \hat H^{(M)}_I\hat S^\dagger &= \sum^2_{i=1} \left[\lambda \chi(\tau)\hat\mu(\tau)\hat\phi(\textbf{x}_i(\tau))\right] \otimes \ket{X_i}_C\bra{X_i}\nonumber \\
    & = \sum^2_{i=1} \hat H^{(D)}_i(\tau) \otimes \ket{X_i}_C\bra{X_i} \nonumber \\
    & = \hat H^{(D)}_I \, ,
\end{align}
and we indeed recover the Hamiltonian for the scenario in which the detector is in a superposition of positions.

The total time-evolution operator in the new frame is given by $\hat U^{(D)} = \hat S \hat U^{(M)} \hat S^\dagger$. Given the static nature of $\hat S$ relative to $\tau$ and the structure of the transformed Hamiltonian, the time-ordered exponential distributes over the sum of orthogonal projectors: 
\begin{align}
    \hat U^{(D)} &= \mathcal{\hat{T}} \exp \left(-i\int \hat H^{(D)}_I(\tau)d\tau\right)\\
    &=\sum^2_{i=1} \ket{X_i}_C\bra{X_i}\otimes \mathcal{T} \exp\left(-i\int\hat H^{(D)}_i(\tau)d\tau\right) \, .
\end{align}
which is precisely the evolution operator for a detector in a superposition of trajectories on a single, fixed background. 

Collecting these results, we obtain the following correspondence. Let $A$ be the frame in which the mass is in a superposition of positions and $B$ the frame in which the detector is in a superposition of positions. The initial state $\hat{\rho}_{i,M}$ in frame $A$ is mapped by the change of quantum reference frame to the initial state $\hat{\rho}_{i,D}$ in frame $B$. In both frames, the detector is initially in its ground state $\ket{0}_D$. In frame $A$, the evolution of the internal state of the detector is controlled by the position of the mass, represented by the states $\ket{M_i}$, whereas in frame $B$ it is controlled by the position of the detector, represented by the states $\ket{X_i}$. The detector dynamics in the $i$-th branch is the same in the two frames. This reflects the fact that the relative position of the detector with respect to the mass is preserved by the transformation and that the state of the field is associated with the position of the mass, so that the detector experiences the same field fluctuations in the corresponding branches. Consequently, after a given interval of proper time, the joint detector-field state is the same in both frames, under the identification of the corresponding field correlation functions.

If in frame $M$ the control is then measured in the basis $\left(\ket{M_1}\pm\ket{M_2}\right)/\sqrt{2}$ and the field degrees of freedom are traced out, the detector excitation probability takes the form given in Eq.~\eqref{eq:transition-pe} with the response functions $\mathcal{F}_{ij}$ described in Eqs.~\eqref{eq:fijM} and \eqref{eq:wightmann-cross-branch}. Likewise, if in frame $D$ the control is measured in the basis $\left(\ket{X_1}\pm\ket{X_2}\right)/\sqrt{2}$ and the field degrees of freedom are traced out, the corresponding excitation probability is given by Eqs.~\eqref{eq:transition-pe} and \eqref{eq:response_fn_ij_def}. To ensure agreement between the probabilities in the two frames, we introduce the correspondence
\begin{equation} \label{eq:map-weightmans}
\mathcal{W}_{ij}^{(M)}(x(\tau_2),x(\tau_1)) = \mathcal{W}^{(D)}_{ij}\left(x_i(\tau_2), x_j\left(\tau_1\right)\right) \, ,
\end{equation}
which defines the cross-branch correlation functions involving vacuum states associated with different geometries in terms of ordinary Wightman functions for the vacuum state on a definite geometry:
\begin{equation}
    \bra{0_i}\hat \phi_i(x(\tau_2))\hat\phi_j(x(\tau_1)) \ket{0_j} = \bra{0}\hat\phi(x_i(\tau_2))\hat\phi(x_j(\tau_1)) \ket{0} \, .
\end{equation}
Under this identification, the cross-branch contribution $\mathcal{F}_{12}$ to the transition probability is identical in the two descriptions, establishing a correspondence between a superposition of geometries and a superposition of trajectories. For brevity, in view of Eq.~\eqref{eq:map-weightmans}, we put
\begin{align}
\mathcal{W}_{ij}(\tau_2,\tau_1) &\equiv \mathcal{W}_{ij}^{(M)}(x(\tau_2),x(\tau_1)) \nonumber \\
&= \mathcal{W}^{(D)}_{ij}\left(x_i(\tau_2), x_j\left(\tau_1\right)\right) \, .
\end{align}

\section{RESPONSE FUNCTION FOR A SUPERPOSITION OF MASSIVE SPHERICAL SHELLS}\label{sec: response_function}

In the previous section, assuming the invariance of physical observables under changes of quantum reference frame, we applied the transformation $\hat{S}$ introduced in Eq.~\eqref{eq:qrf_transformation_def} to define correlation functions for a scalar field evaluated at events along the worldline of a detector moving in a superposition of two geometries. These functions were obtained from the corresponding correlation functions in a frame with a definite background geometry in which the detector is in a superposition of two worldlines.

We now consider the concrete case of a superposition of spacetimes produced by a massive, transparent shell of mass $M$ and radius $R$, noninteracting with the scalar field, prepared in a superposition of two positions. The Wightman functions can be computed in the frame in which the geometry is definite. The case of a UDW detector at a fixed position at the center of the shell was studied in~\cite{Ng:2016hzn}. Here, we extend this analysis to a UDW detector in a superposition of two generic positions inside the shell. We follow the presentation of~\cite{Ng:2016hzn} for the description of the normal modes of the scalar field in the spacetime generated by the massive shell.

The geometry produced by a massive shell at a definite position is a Schwarzschild exterior glued to a flat interior at the shell. The metric can be written as 
\begin{equation}\label{eq:metric_def}
\mathrm{d} s^2= \begin{cases}-f(r) \mathrm{d} t^2+\frac{1}{f(r)} \mathrm{d} r^2+r^2 \mathrm{~d} \Omega^2, & r>R \\ -f(R) \mathrm{d} t^2+\mathrm{d} r^2+r^2 \mathrm{~d} \Omega^2, & r<R,\end{cases}
\end{equation}
where $d \Omega^2=d \theta^2+\sin ^2 \theta \, d \varphi^2$ is the metric of the unit $2$-sphere and
\begin{equation}
    f(r)=1-\frac{2M}{r} \, .
\end{equation}
The constant factor $f(R)$ in the time component of the interior metric is required for the line element in Eq.~\eqref{eq:metric_def} to satisfy the first junction condition \cite{Israel:1966rt,Poisson:2009pwt}: the induced metric must be continuous across the shell, which is essential for the glued spacetime to be a well-defined solution to Einstein's equations. Physically, the factor $f(R)$ represents the constant gravitational redshift of the clocks inside the shell relative to infinity. The radial component $g_{rr}$ is not continuous and has a jump discontinuity across the massive shell.

For a static detector inside the shell, the relation between the coordinate time $t$ and the detector's proper time $\tau$ is $t=\tau/\alpha$, where
\begin{equation}
    \alpha=\sqrt{f(R)} \, .
\end{equation}
We consider a detector in a superposition of two fixed positions, both in the interior of the shell. Since the detector is static in both branches, the coordinate time along each worldline satisfies $t_i(\tau)=\tau/\alpha$, $i=1,2$. The corresponding worldlines can therefore be written as
\begin{equation}
    x_i(\tau)
    =
    \left(
    \frac{\tau}{\alpha},
    r_i,\theta_i,\varphi_i
    \right),
    \quad
    0<r_i<R \, .
\end{equation}

The detector interacts with the scalar field $\hat{\phi}(x)$, which satisfies the massless Klein-Gordon equation,
\begin{equation}\label{eq:KG_def}
    \Box \hat{\phi} =\frac{1}{\sqrt{-g}}\partial_\mu\left(\sqrt{-g}\, g^{\mu\nu}\partial_\nu \hat{\phi}\right)=0 \, .
\end{equation}
The field operator can be expanded in normal modes as
\begin{equation}\label{eq:phi_mode_expansion_def}
    \hat{\phi}(x) = \sum_{l, m} \int_0^{\infty} \mathrm{d} \omega\left(\hat{a}_{\omega l m} \Psi_{\omega l m}(x)+\hat{a}_{\omega l m}^{\dagger} \Psi_{\omega l m}^{\dagger}(x)\right) \, ,
\end{equation}
where the modes $\Psi_{\omega l m}$ are solutions of the Klein-Gordon equation \eqref{eq:KG_def} with frequency $\omega$ (with respect to the time coordinate $t$) and angular momentum numbers $l,m$, while $\hat{a}_{\omega l m}$ and $\hat{a}^\dagger_{\omega l m}$ are the corresponding mode annihilation and creation operators satisfying canonical commutation relations
\begin{align}
[\hat{a}_{\omega' l' m'}, \hat{a}_{\omega lm}^{\dagger}] &= \delta(\omega' - \omega) \delta_{l' l} \delta_{m^\prime m} \, ,
\nonumber \\
[\hat{a}_{\omega' l' m'}, \hat{a}_{\omega lm}] &= [\hat{a}_{\omega' l' m'}^{\dagger}, \hat{a}_{\omega lm}^{\dagger}] =0 \, .
\end{align}
Due to spherical symmetry, the modes admit separable solutions of the form
\begin{equation}
   \Psi_{\omega l m} = \frac{1}{\sqrt{4\pi\omega}} e^{-i\omega t}  \psi_{\omega l}(r) Y_{l m}(\theta,\phi). 
\end{equation}
where $Y_{lm}(\theta,\phi)$ are spherical harmonics and $\psi_{\omega l}(r)$ are the radial modes.

The Wightman function defined in Eq.~\eqref{eq:wightman_fn_ij_def}, evaluated along the detector trajectories, is now given by the mode sum
\begin{align}
    \mathcal{W}_{ij}(\tau_2, \tau_1) &= \mathcal{W}_{ij}^{(D)}\left(x_i(\tau_2), x_j\left(\tau_1\right)\right)
    \nonumber \\
    &=\sum_{ l,m} \int_0^\infty d\omega\, \Psi_{\omega l m}(x_i(\tau_2)) \Psi_{\omega l m}^{*}(x_j(\tau_1)).
\end{align}
Using the explicit form of the modes, this becomes
\begin{multline} \label{eq:Wij-mode-sum}
    \sum_{ l,m} \int_0^\infty \frac{d\omega}{4\pi\omega}e^{-i\omega(\tau_2-\tau_1)/\alpha}\psi_{\omega l}(r_i)\psi_{\omega l}^{*}(r_j)
    \\
    \times Y_{ l m}(\theta_i,\varphi_i)Y_{ l m}^{*}(\theta_j,\varphi_j).
\end{multline}
When both radii are nonzero, $r_i > 0$, the unit vectors $\hat r_i$ are well defined, and the angular dependence simplifies due to the addition theorem for the spherical harmonics,
\begin{equation}
    \sum_{m=- l}^{ l}Y_{ l m}(\theta_i,\varphi_i)Y_{ l m}^{*}(\theta_j,\varphi_j)=\frac{2 l+1}{4\pi}P_l(\cos\gamma_{ij}),
\end{equation}
where $\gamma_{ij}$ is the angular separation between the two positions,
\begin{align}
    \cos\gamma_{ij}&=\hat r_i \cdot \hat r_j
    \nonumber \\
    &=\cos\theta_i\cos\theta_j+\sin\theta_i\sin\theta_j\cos(\varphi_i-\varphi_j).
\end{align}
Thus, Eq.~\eqref{eq:Wij-mode-sum} can be written as
\begin{multline}
    \mathcal{W}_{ij}(\tau_2,\tau_1)=\sum_{ l=0}^{\infty}\int_0^\infty\frac{d\omega}{4\pi\omega}e^{-i\omega(\tau_2-\tau_1)/\alpha}\psi_{\omega l}(r_i)\psi_{\omega l}^{*}(r_j)\\
    \times\frac{2 l+1}{4\pi}P_l(\cos\gamma_{ij})\, .
    \label{eq:Wij-addition}
\end{multline}

As shown in~\cite{Ng:2016hzn}, the regular interior solutions of the Klein--Gordon equation inside the shell are given by spherical Bessel functions,
\begin{equation} \label{eq:normalization-A}
\psi_{\omega l}(r)=A_{\omega l}\,j_l(\tilde{\omega}r) \, ,
\end{equation}
where $\tilde{\omega}=\omega/\alpha$, and $A_{\omega l}$ is a normalization constant. 
For completeness, we describe the derivation of the normal modes and their normalization in Appendix~\ref{appendix:radial_modes}.

Changing the integration variable from $\omega$ to $\tilde{\omega}$, the Wightman function becomes
\begin{align}
    \mathcal{W}_{ij}(\tau_2,\tau_1)=&\int_0^\infty\frac{d\tilde{\omega}}{16\pi^2\tilde{\omega}}e^{-i\tilde{\omega}(\tau_2-\tau_1)}\mathcal{M}_{ij},
\end{align}
where 
\begin{equation}
    \mathcal{M}_{ij}=\sum_{ l=0}^{\infty} (2 l+1)|A_{\omega l}|^2j_{l}(\tilde{\omega}r_i)j_{l}(\tilde{\omega}r_j)P_l(\cos\gamma_{ij}) \, .
\label{eq:sum_l}
\end{equation}
For real radial modes, one has
\begin{equation}
    \mathcal{W}_{21}(\tau_2,\tau_1)=\mathcal{W}_{12}(\tau_2,\tau_1).
\end{equation}
Substituting the Wightman function expressed as a sum over modes into Eq.~\eqref{eq:response_fn_ij_def}, we obtain
\begin{multline}
    \mathcal{F}_{ij}(\Omega)=\int_{-\infty}^{\infty}d\tau_1\int_{-\infty}^{\infty}d\tau_2~\chi(\tau_1)\chi(\tau_2)~e^{-i\Omega(\tau_2-\tau_1)}\\
    \times\int_0^\infty\frac{d\tilde{\omega}}{16\pi^2\tilde{\omega}}e^{-i\tilde{\omega}(\tau_2-\tau_1)} \mathcal{M}_{ij}\nonumber.
\end{multline}
The double integral over proper time factorizes as a product of independent integrals, each reducing to a Fourier transform of the switching function. Using the following definition for the Fourier transform, 
\begin{equation}
    \hat{\chi}(\Omega)=\frac{1}{\sqrt{2\pi}}\int_{-\infty}^{\infty}d\tau\,e^{-i\Omega\tau}\chi(\tau),
\end{equation}
we obtain
\begin{equation}
     \mathcal{F}_{ij}(\Omega)=\int_0^\infty \frac{d\tilde{\omega}}{8\pi\tilde{\omega}}\left|\hat{\chi}(\Omega+\tilde{\omega})\right|^2 \mathcal{M}_{ij} .
\end{equation}
Now we can write down the total response function of the detector as: 
\begin{align}\label{eq:response_fn_final_form}
    \mathcal{F}_{\pm}(\Omega)&=\mathcal{F}_{11}+\mathcal{F}_{22}\pm2\mathcal{F}_{12} \nonumber\\
    &=\int_0^\infty\frac{d\tilde{\omega}}{8\pi\tilde{\omega}}\left|\hat{\chi}(\Omega+\tilde{\omega})\right|^2\left[ \mathcal{M}_{11}+\mathcal{M}_{22}\pm2\mathcal{M}_{12}\right] .
\end{align}

\begin{table}
\caption{\label{tab:Mij}
Expressions for $\mathcal{M}_{ij}$ for a detector in a superposition
of two positions inside the shell.}
\begin{ruledtabular}
\begin{tabular}{c|l}

\rule[-1.2ex]{0pt}{5ex}
$\mathcal{M}_{ij}$ & One position centered $(r_1=0,\ 0<r_2<R)$\\[2ex]
\colrule

\rule[-1.2ex]{0pt}{5ex}
$\mathcal{M}_{11}$ &
$|A_{\omega 0}|^2$ \\[2pt]

\rule[-1.2ex]{0pt}{5ex}
$\mathcal{M}_{22}$ &
$\displaystyle\sum_l(2l+1)
 |A_{\omega l}j_l(\tilde\omega r_2)|^2$ \\[2pt]

\rule[-1.2ex]{0pt}{5ex}
$\mathcal{M}_{12}=\mathcal{M}_{21}$ &
$|A_{\omega 0}|^2j_0(\tilde\omega r_2)$ \\[2pt]

\colrule
\rule[-1.2ex]{0pt}{5ex}
$\mathcal{M}_{ij}$ & Both off-center $(0<r_i<R)$ \\[2ex]
\colrule

\rule[-1.2ex]{0pt}{5ex}
$\mathcal{M}_{11}$ &
$\displaystyle\sum_l(2l+1)
 |A_{\omega l}j_l(\tilde\omega r_1)|^2$ \\[2pt]

\rule[-1.2ex]{0pt}{5ex}
$\mathcal{M}_{22}$ &
$\displaystyle\sum_l(2l+1)
 |A_{\omega l}j_l(\tilde\omega r_2)|^2$ \\[2pt]

\rule[-1.2ex]{0pt}{5ex}
$\mathcal{M}_{12}=\mathcal{M}_{21}$ &
$\begin{aligned}[t]
&\displaystyle\sum_l(2l+1)|A_{\omega l}|^2
 j_l(\tilde\omega r_1)j_l(\tilde\omega r_2)
 P_l(\cos\gamma_{12})
\end{aligned}$

\end{tabular}
\end{ruledtabular}
\end{table}

In Table \ref{tab:Mij}, we give the explicit expressions for the components $\mathcal{M}_{ij}$ for two different configurations of the detector positions. If the detector is at the center of the shell in either branch of the superposition, then the behavior of the spherical Bessel functions at the origin implies $j_l(0)=\delta_{ l,0}$, and hence $ \psi_{\omega l}(0)=A_{\omega l}\,\delta_{ l,0}$. Therefore, the Wightman functions receive contributions only from the $l=0$ mode whenever either of their arguments lies at the center. The same-branch Wightman function receives contributions from all angular momenta for an off-center detector. The diagonal term $\mathcal{M}_{22}$ is identical in both blocks since the detector is off-center in branch 2 in both configurations.

\section{Numerical results}\label{sec:numerics}

We now evaluate the response function of the UDW detector in the quantum reference frame in which the shell is at a definite position and the detector in a superposition of two positions. We denote by $\mathbf r_i$ the position of the detector in the $i$-th branch, and by $\gamma$ the angle between the position vectors in the two branches, $\cos \gamma = \mathbf r_1 \cdot \mathbf r_2/(r_1 r_2)$. We consider a broad class of configurations for the two possible positions of the detector. The numerical calculations are performed for a shell of mass $M=1/2$ and radius $R=3 R_{Sch}$, where $R_{Sch}=2M$ is the Schwarzschild radius associated with the mass of the shell. We use natural units in which $G=c=\hbar=1$.

The detector is switched on with a Gaussian profile $\chi(\tau)=\exp(-\tau^2/2\sigma^2)$ of width $\sigma=1/2$. The normalization constants $A_{\omega l}$ entering Eq.~\eqref{eq:normalization-A} for the radial modes are precomputed for a range of $\{\omega,l\}$ values and interpolated for efficient computation. The sum over $l$ in the Eq.~\eqref{eq:sum_l} for $M_{ij}$ is truncated at a cutoff $l_{\max}$ determined by the properties of the spherical Bessel functions $j_l(\tilde\omega r)$. For a fixed argument $x=\tilde{\omega}r$, $j_l(x)$ is strongly suppressed when $l\gg x$. Accordingly, for each value of $\tilde{\omega}$, we choose $l_{max}$ sufficiently larger than $\tilde{\omega}r_{max}$, with $r_{max}=\max(r_i,r_j)$, so that the neglected terms of the sum give a negligible contribution to $F_{ij}$.

With the numerical setup in place, we now study how the response function depends on the relative placement of the two detector positions in superposition, both taken inside the shell. Our focus is on the post-selected response functions $\mathcal{F}_{\pm}(\Omega)$, given in Eq.~\eqref{eq:response_fn_final_form}, together with the cross contribution $\mathcal{F}_{12}$. We first compute these quantities in the presence of the shell, and then compare the results with those in flat spacetime with the same separation between the two detector positions, to isolate the contribution due to the presence of the shell.

\begin{figure}
    \centering
    \subfloat[]{
        \includegraphics[width=\columnwidth]{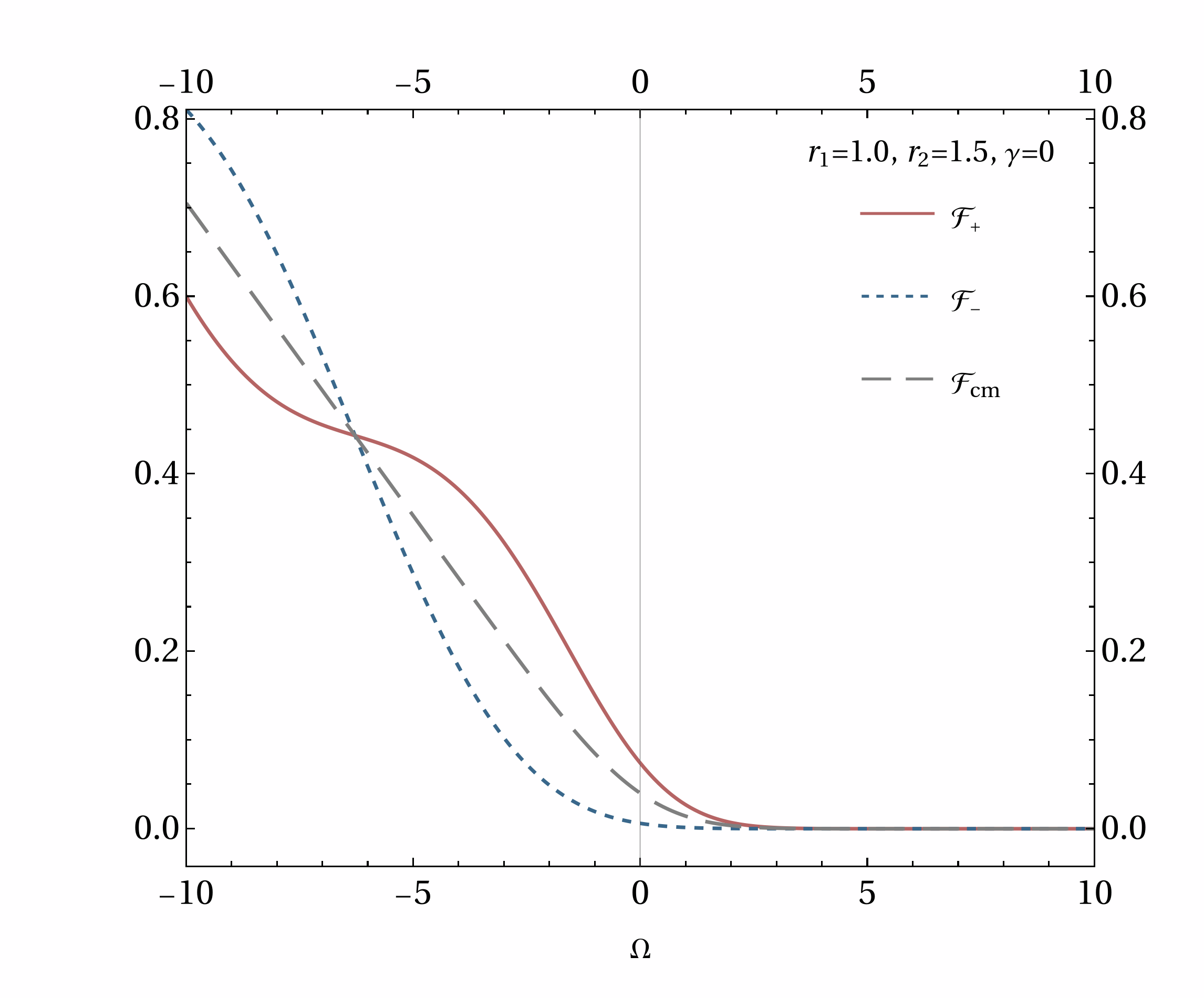}
        \label{fig:img1}
    }
    \vspace{0.5cm}
    \subfloat[]{
        \includegraphics[width=0.40\columnwidth]{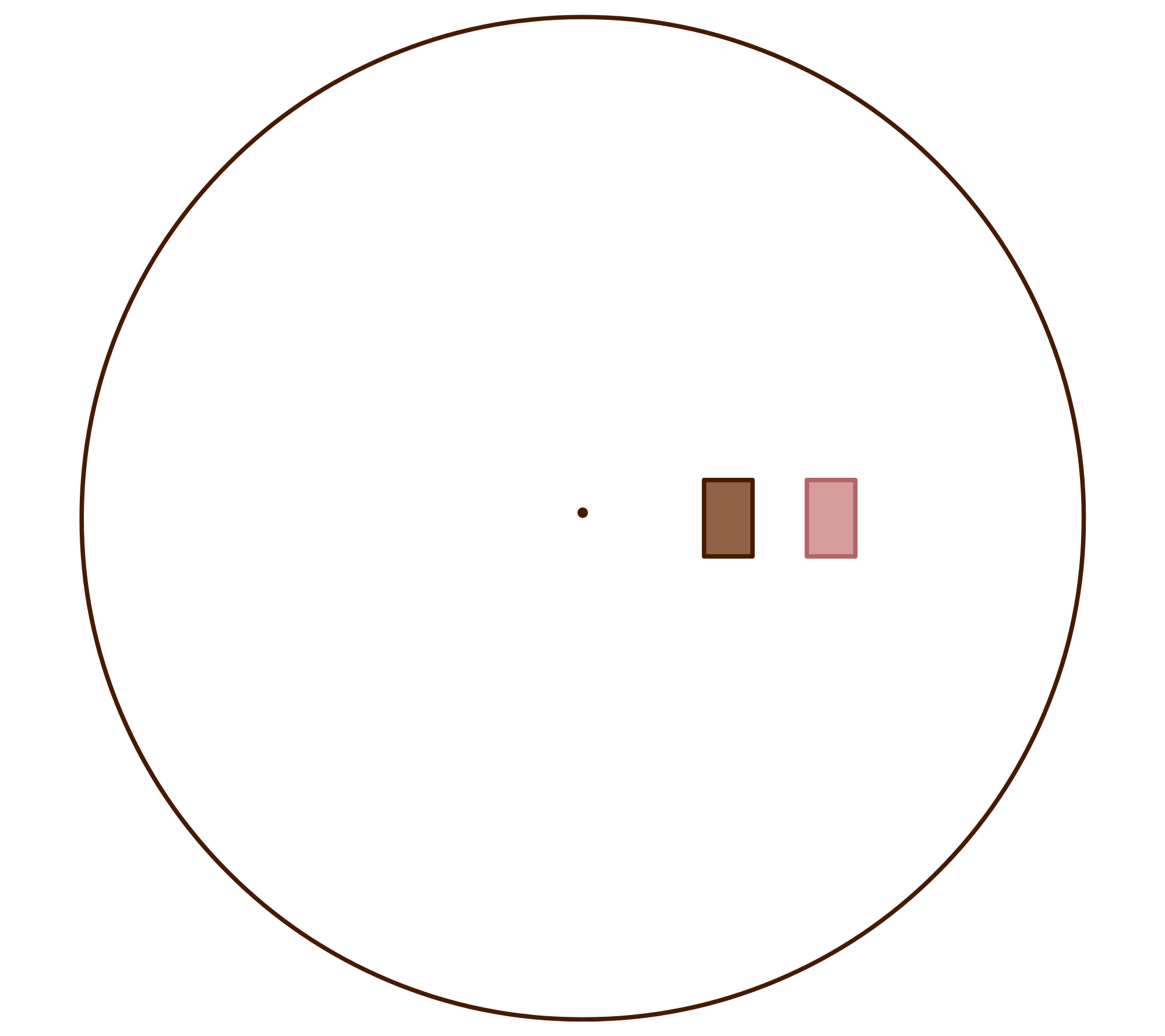}
        \label{fig:img2}
    }
    \hfill
    \subfloat[]{
        \includegraphics[width=0.40\columnwidth]{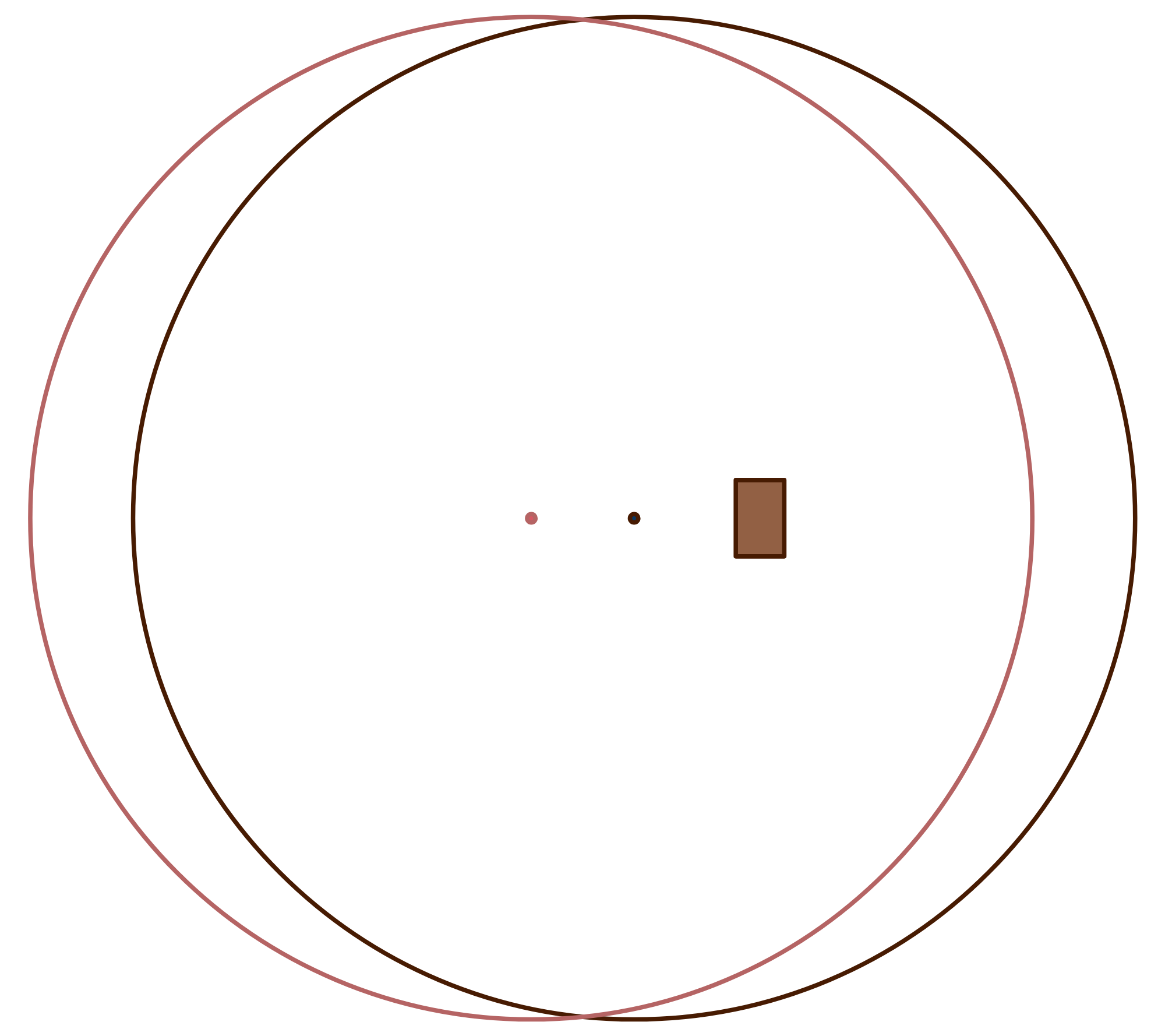}
        \label{fig:img3}
    }
    \caption{Total detector response $\mathcal{F}_{\pm}(\Omega)$ for a spatial superposition of two fixed positions $r_1 = 1.0$ and $r_2 = 1.5$. The solid red and dashed blue curves represent the post-selected responses $\mathcal{F}_{+}$ and $\mathcal{F}_{-}$, respectively, exhibiting interference oscillations. The dashed grey curve describes the classical mixture $\mathcal{F}_{cm}$, which represents the incoherent combination of the two post-selected responses. The circular insets illustrate the spatial configurations: (b) represents the positions of the detector in the quantum reference frame in which it is in a spatial superposition and (c) depicts the shell positions in the frame where the mass is delocalized and the detector at a definite position.}
    \label{fig:C0_tot}
\end{figure}

We start by analyzing general features of the response function for a detector in a superposition of two fixed positions. Figure~\ref{fig:C0_tot}
 shows the post-selected responses $\mathcal{F}_{\pm}(\Omega)$ alongside the classical mixture $\mathcal{F}_{cm} = (\mathcal{F}_{11} + \mathcal{F}_{22})/4$, which corresponds to the incoherent combination of the response from the two branches. The classical mixture exhibits a smooth monotonic behavior similar to that of a single detector inside a shell. In contrast, $\mathcal{F}_{\pm}$ show characteristic oscillations about $\mathcal{F}_{cm}$ that are a direct consequence of the cross-correlation term $\mathcal{F}_{12}$, which encodes the quantum coherence of the superposition. A more detailed analysis of $\mathcal{F}_{12}$ for different detector configurations is presented below.

\begin{figure}
    \centering
    \includegraphics[width=\linewidth]{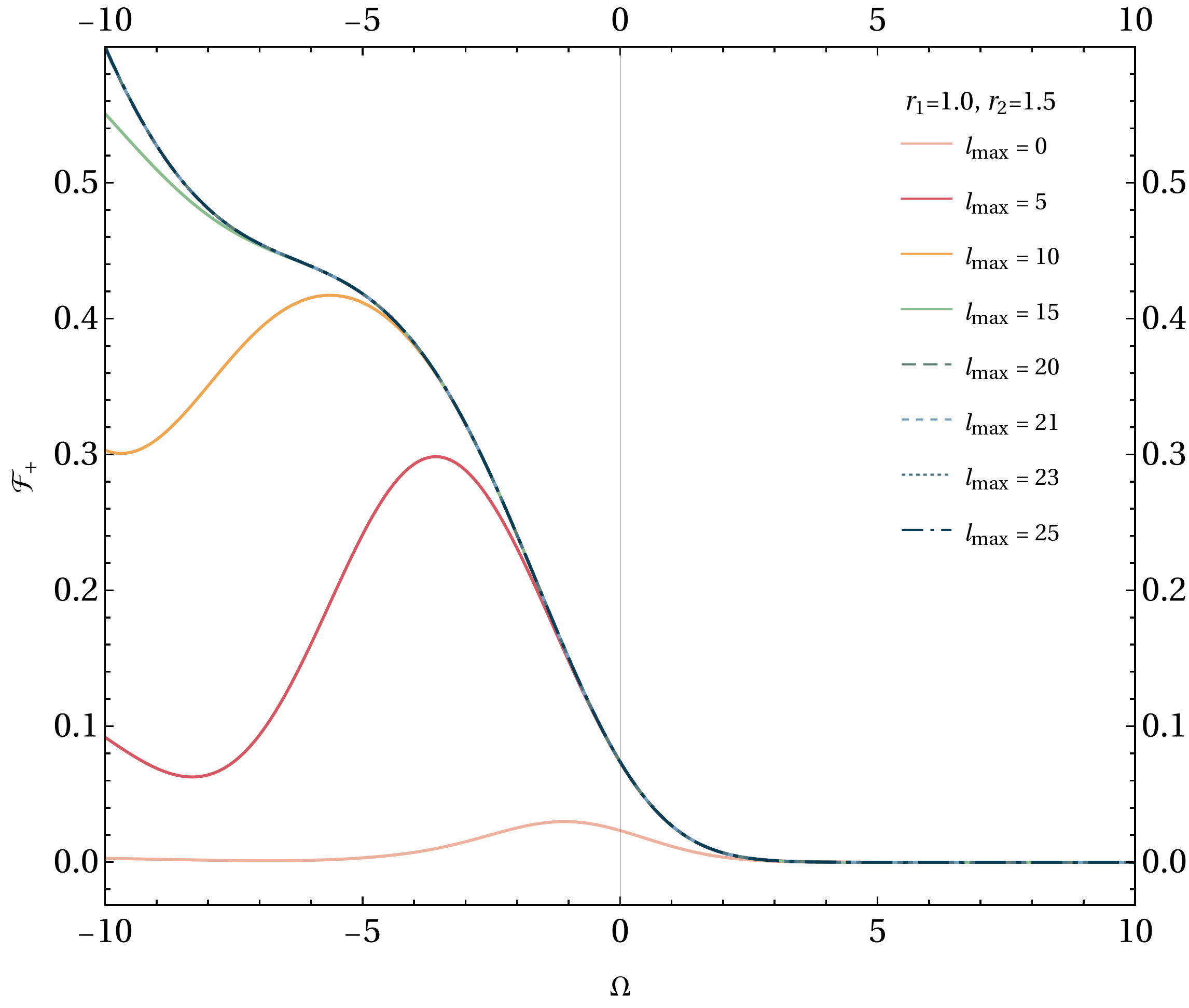}
    \caption{Numerical convergence of the response function as a function of the cutoff $l_{max}$. A configuration where both detector branches are off-center ($r_1 = 1, r_2 = 1.5)$ is considered. As $l_{max}$ increases, the curves stabilize.  Convergence is achieved at $l_{max} \approx 20$ for this setup and window of energy gaps.}
    \label{fig:C0_convergence}
\end{figure}

Given the numerical approximations required for the evaluation of the response function, it is crucial to ensure the robustness of the results and verify that the observed features are physical signatures rather than numerical artifacts.  To this end, we systematically test the numerical parameters for convergence. Figure~\ref{fig:C0_convergence} shows the convergence test for the cutoff $l_{max}$  for a configuration in which both detector positions are off-center. In this case, as discussed in Section~\ref{sec: response_function}, the response function receives appreciable contributions from higher angular momentum modes ($l>0$). Fig.~\ref{fig:C0_convergence} shows that the sum over $l$ in Eq.~\eqref{eq:sum_l} converges as $l_{max}$ increases, supporting the choice of cutoff used in our numerical calculations.

After addressing the issue of convergence, we now probe the dependence of the response function on the geometry of the superposition of positions. We start with a simple class of configurations obtained by keeping the detector at the center of the shell in one branch of the superposition, $r_1=0$, and displacing it radially outward in the other branch. The coordinate distance between the two positions, $D=|\mathbf{r}_2-\mathbf{r}_1| = r_2$, increases as the position in the second branch approaches the shell. The results are shown in Fig.~\ref{fig:Conf1}. The geometry of the setup is depicted in the panels (d1) and (d2) in the figure. Panel~(d1) shows the detector positions in the frame where the detector is delocalized, while panel~(d2) depicts the frame in which the shell is in a superposition of positions.

\manyfiguresb{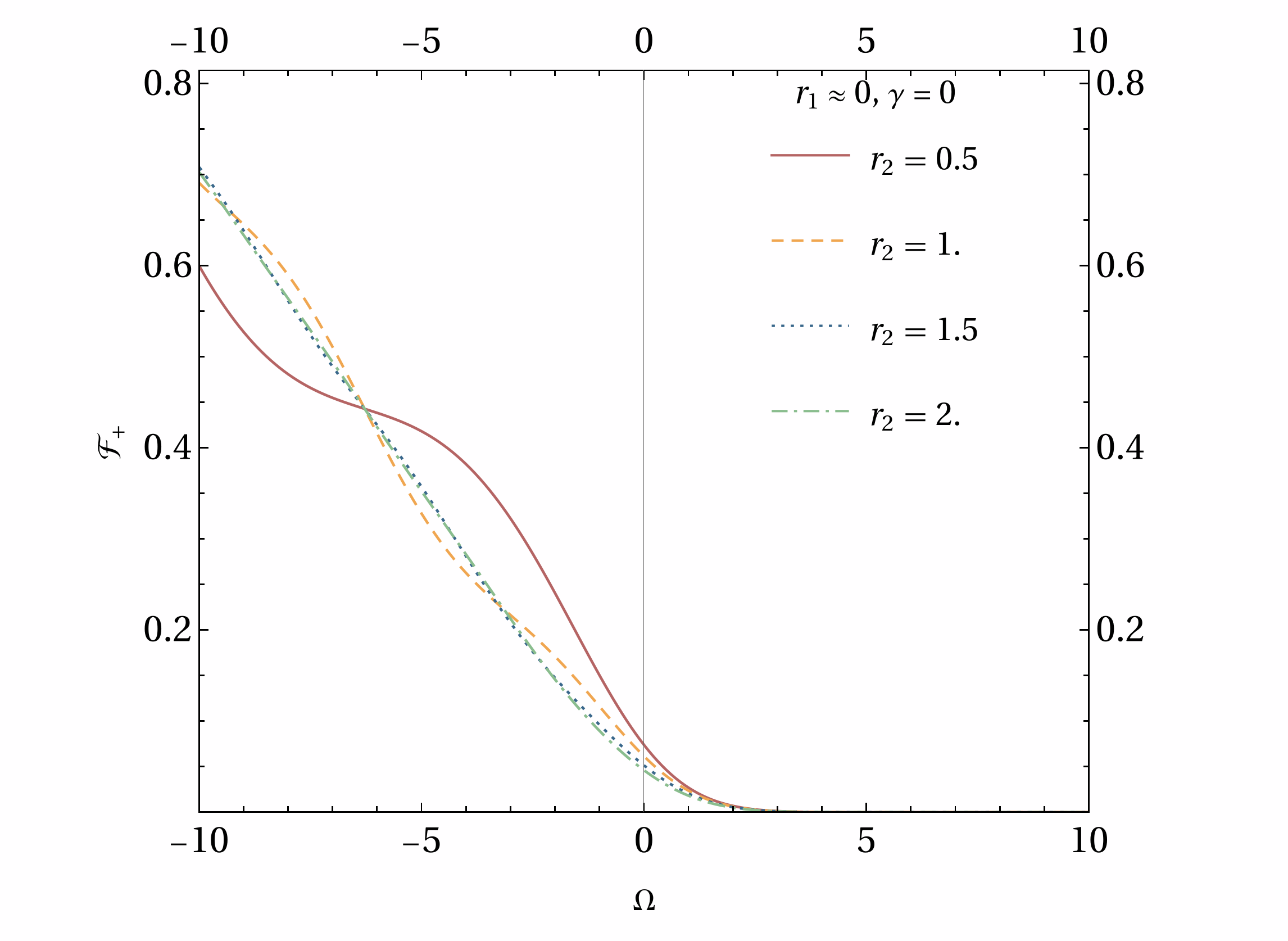}{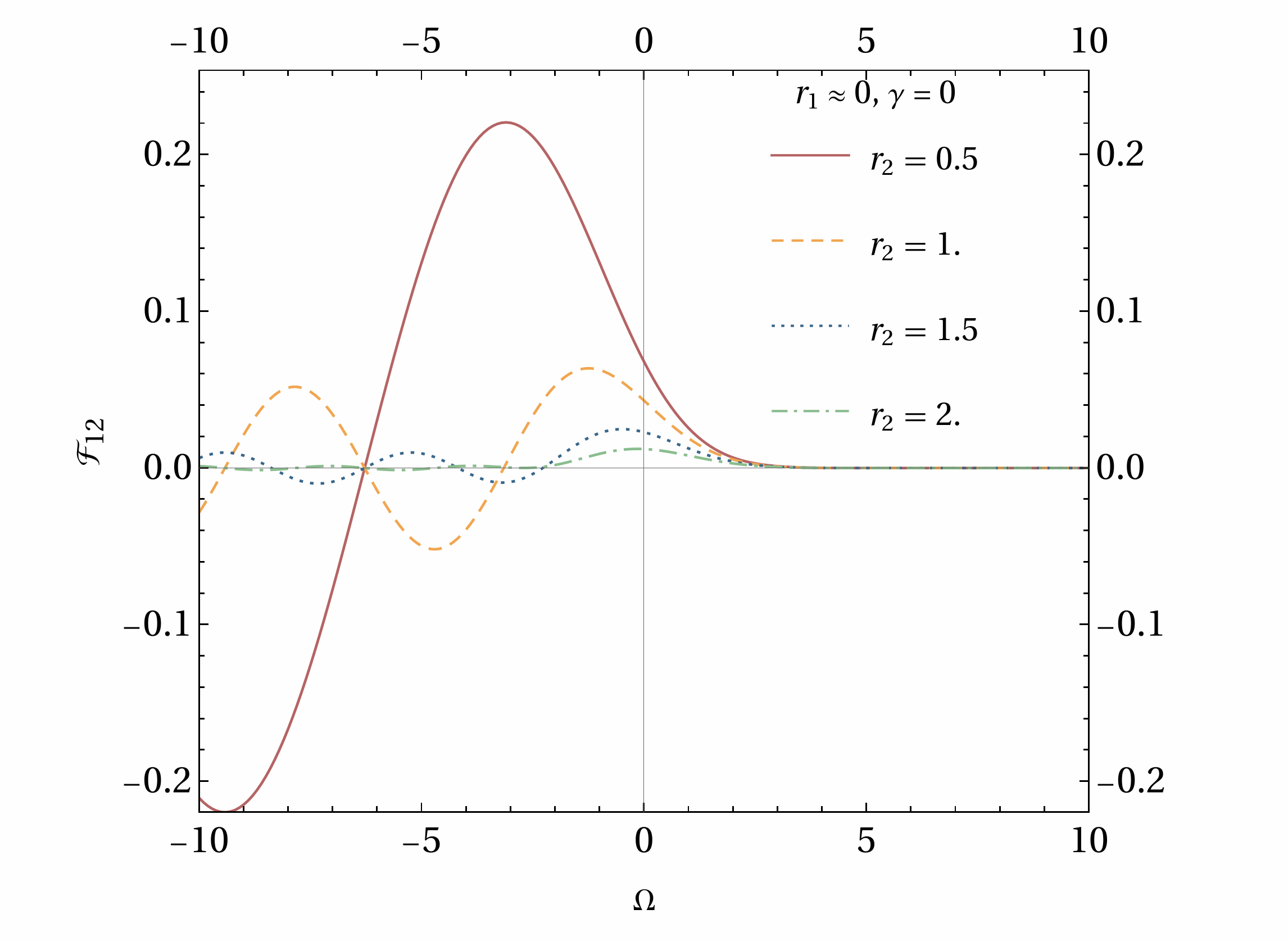}{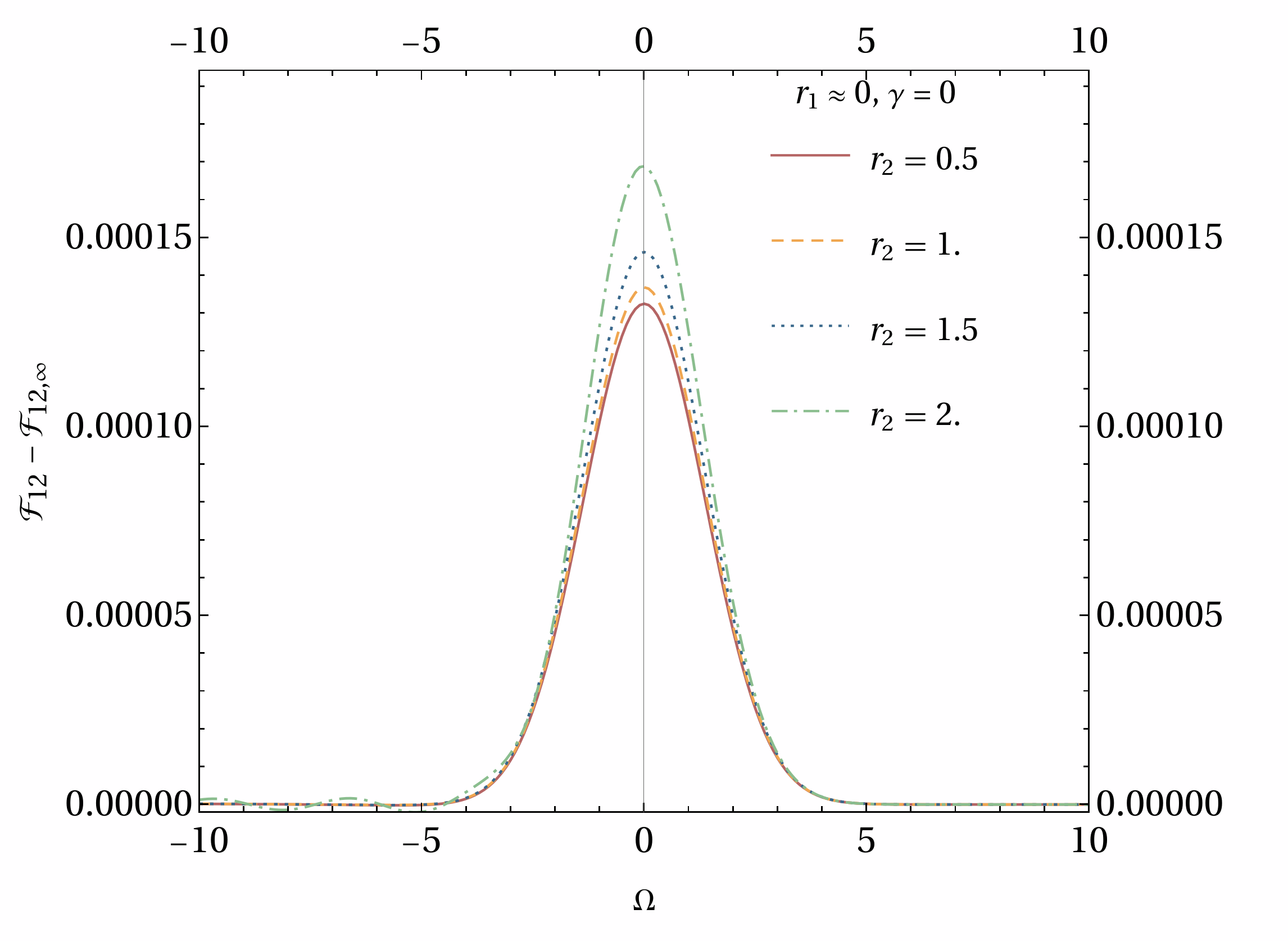}{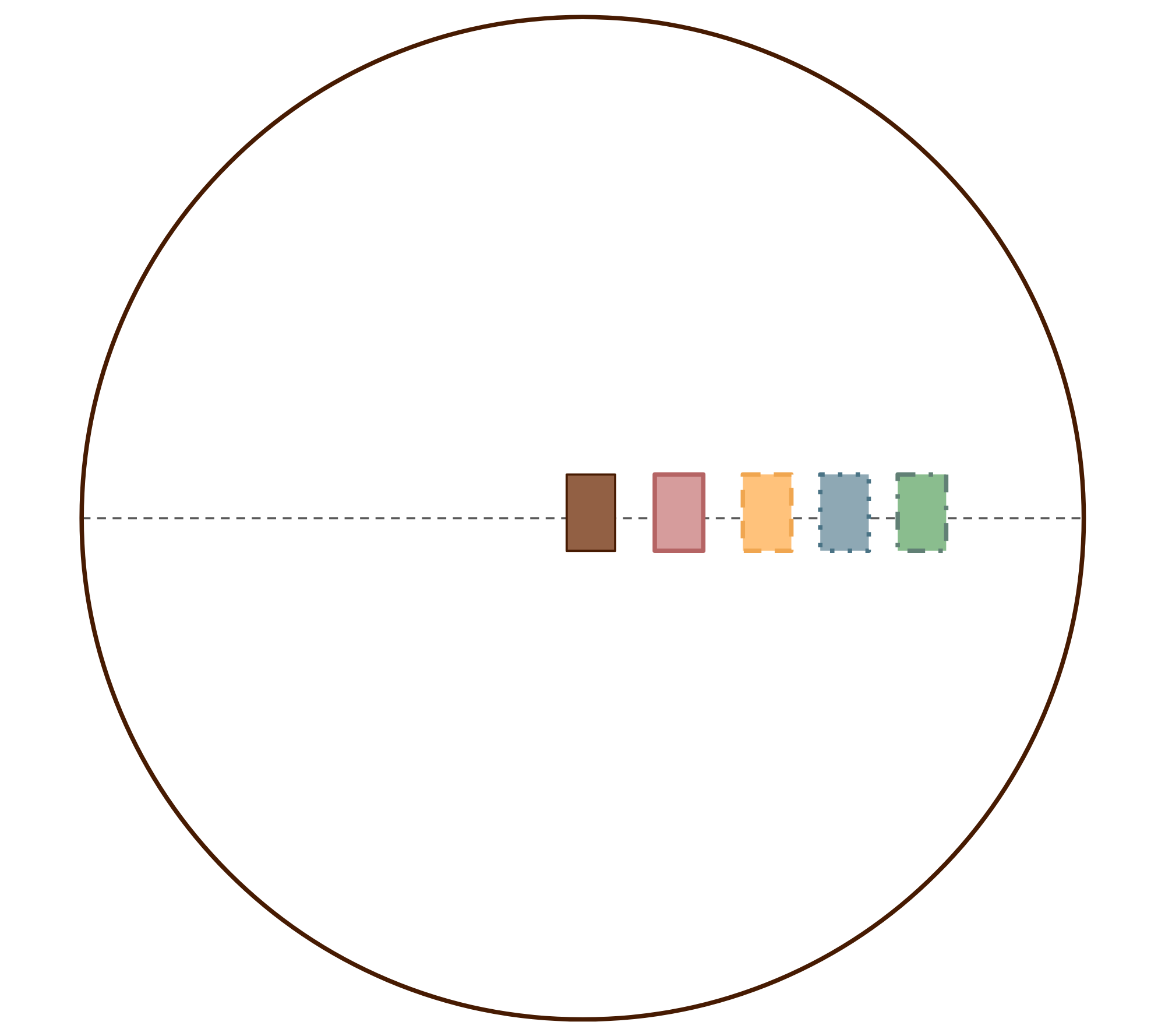}{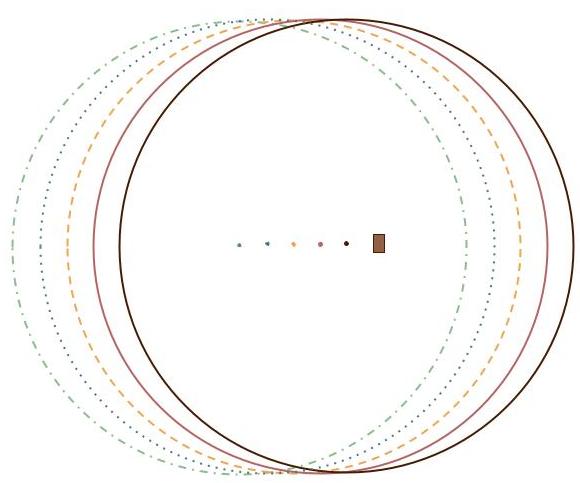}{Numerical results for the configuration with $r_1 = 0$ and $r_2$ displaced along the diameter of the shell. Panels (d1) and (d2) represent the setup in the frame in which the shell is at a fixed position and in the frame in which the shell is in a superposition of positions, respectively. Panel (a) shows the total response functions $\mathcal{F}_{+} (\Omega)$, while panel (b) shows the cross-branch contribution $\mathcal{F}_{12}$. Panel (c) displays the shell-induced correction, defined as the difference between the response function for a shell of radius $R=3$ and that in the limit $R\rightarrow \infty$ for the same detector configuration. }{fig:Conf1}{0 0 0 0}{1\textwidth}

As described in Table~\ref{tab:Mij}, placing one detector position at the center greatly simplifies the interference term $\mathcal{M}_{12}$ and the same-branch term $\mathcal{M}_{11}$, which are then completely determined by the monopole contribution ($l=0$). The off-center same-branch term $\mathcal{M}_{22}$ still receives contributions from all angular momenta. As $r_2$ increases, two main features can be observed in Fig.~\ref{fig:Conf1}(b): the amplitude of $\mathcal{F}_{12}$ decreases and its oscillations in $\Omega$ become faster. Both follow from the dependence of $\mathcal{M}_{12}$ on $\sin(\tilde\omega r_2)/(\tilde\omega r_2)$. This interference pattern is imprinted on the total post-selected response $\mathcal{F}_{+}$ as oscillations about the response corresponding to a classical mixture of the two branches.

To isolate the effect of the shell, we subtract from $\mathcal{F}_{12}$ its value for an infinitely large shell, $\mathcal{F}_{12,\infty}$. As $R\to\infty$, the matching surface recedes to infinity, and the normal modes reduce to those of global Minkowski space. Thus, $\mathcal{F}_{12,\infty}$ provides a flat-spacetime reference obtained within the same mode-sum framework. The difference $\mathcal{F}_{12}-\mathcal{F}_{12,\infty}$ is shown in Fig.~\ref{fig:Conf1}(c). Its magnitude is of order $10^{-4}$, with the maximum difference occurring at an energy gap $\Omega\approx0$. This difference, induced by the presence of the shell, grows with $r_2$: the closer the detector lies to the shell in the off-center branch, the stronger the modification of the cross-correlations between the branches due to the shell.

Let us now analyze the effect of the orientation of the relative position between the detector positions within the shell. To this end, we consider a class of configurations in which the detector is fixed at $r_1=0.5$ along the $x$-axis in the branch $1$, while in branch $2$ its position is varied along a circle of unit radius centered at the detector position in branch 1. The coordinate separation between the two detector positions is then kept fixed, $D=1$, while the radial and angular coordinates of the detector in branch 2 are varied.

\manyfiguresb{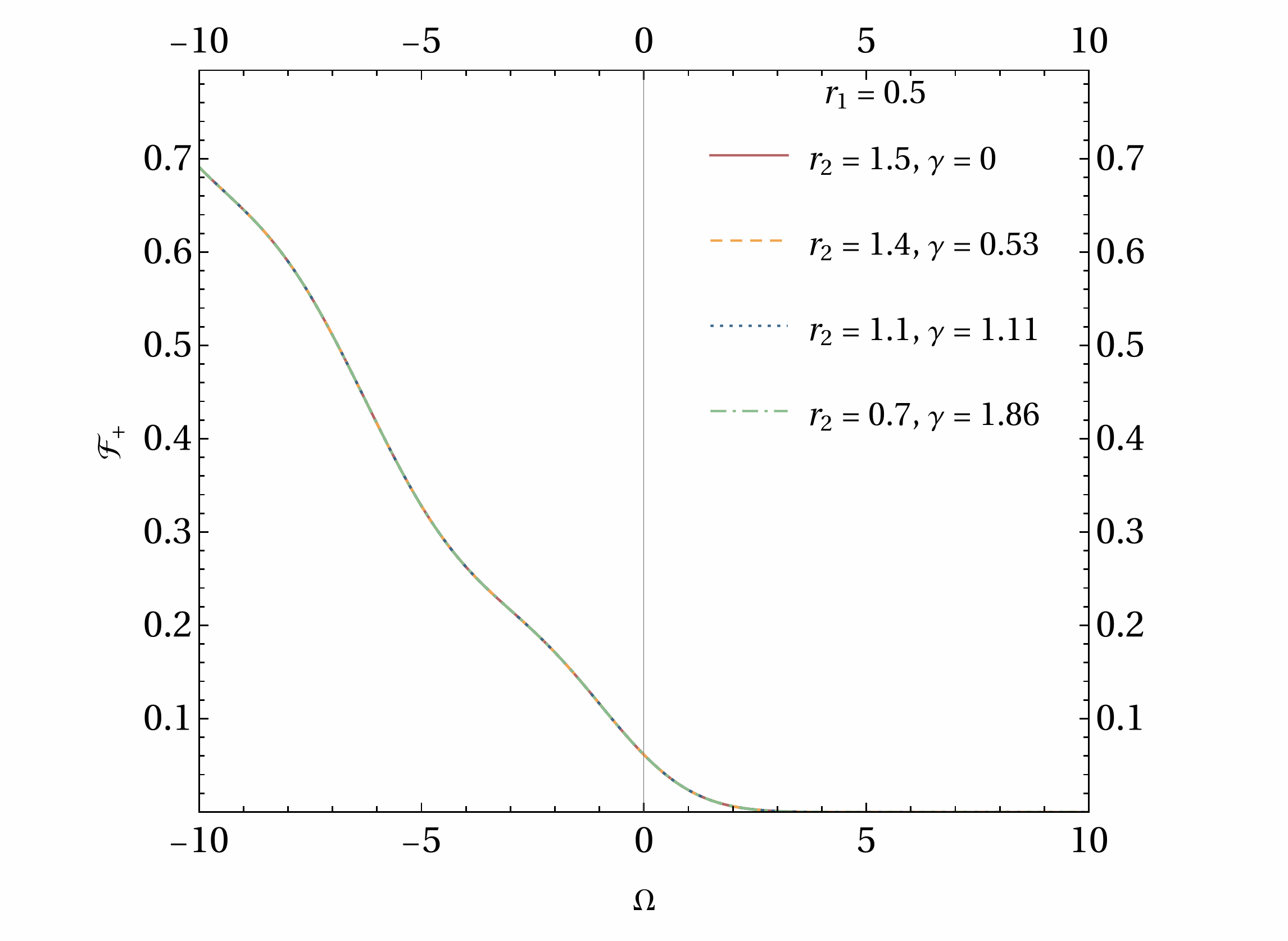}{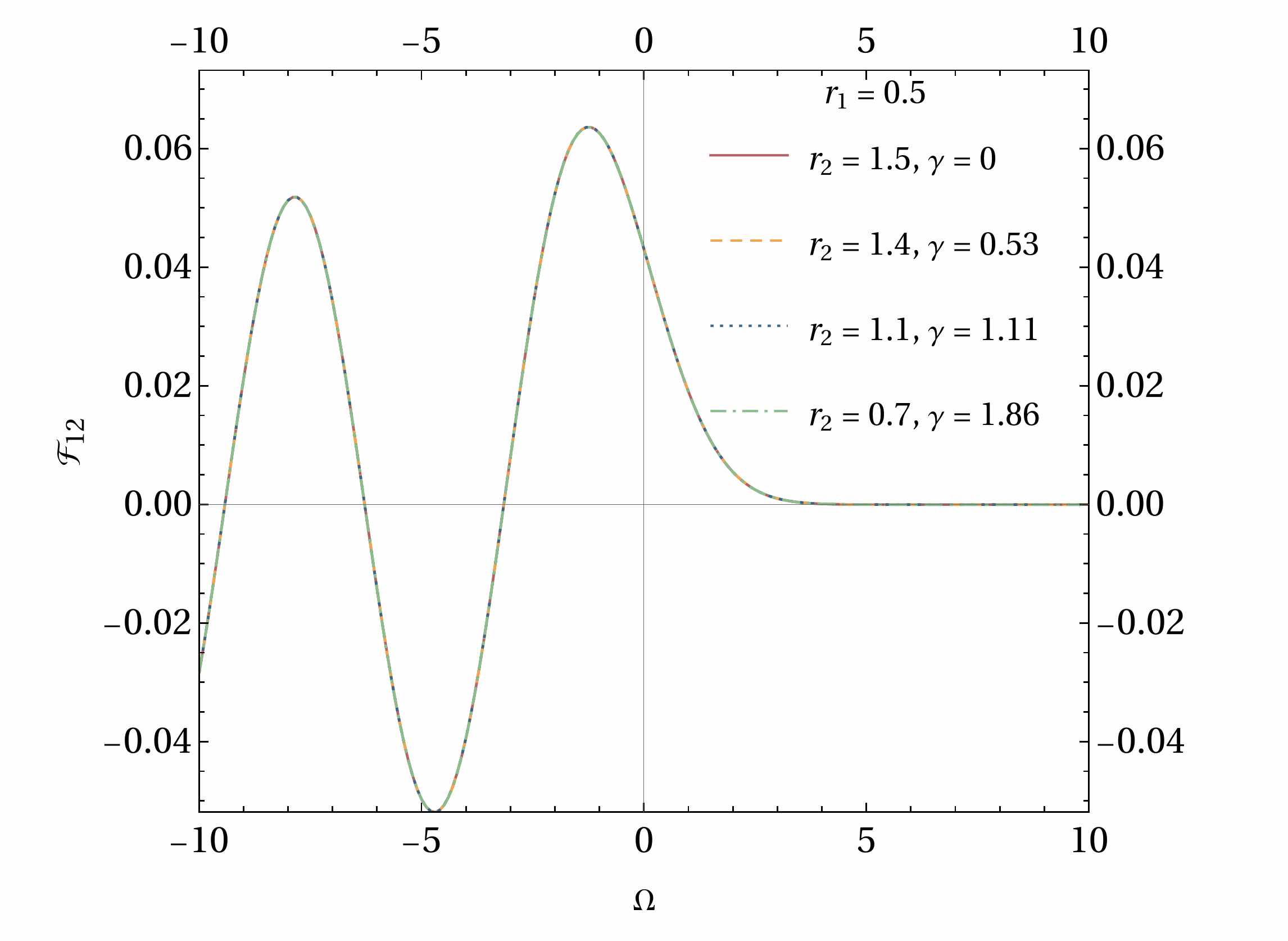}{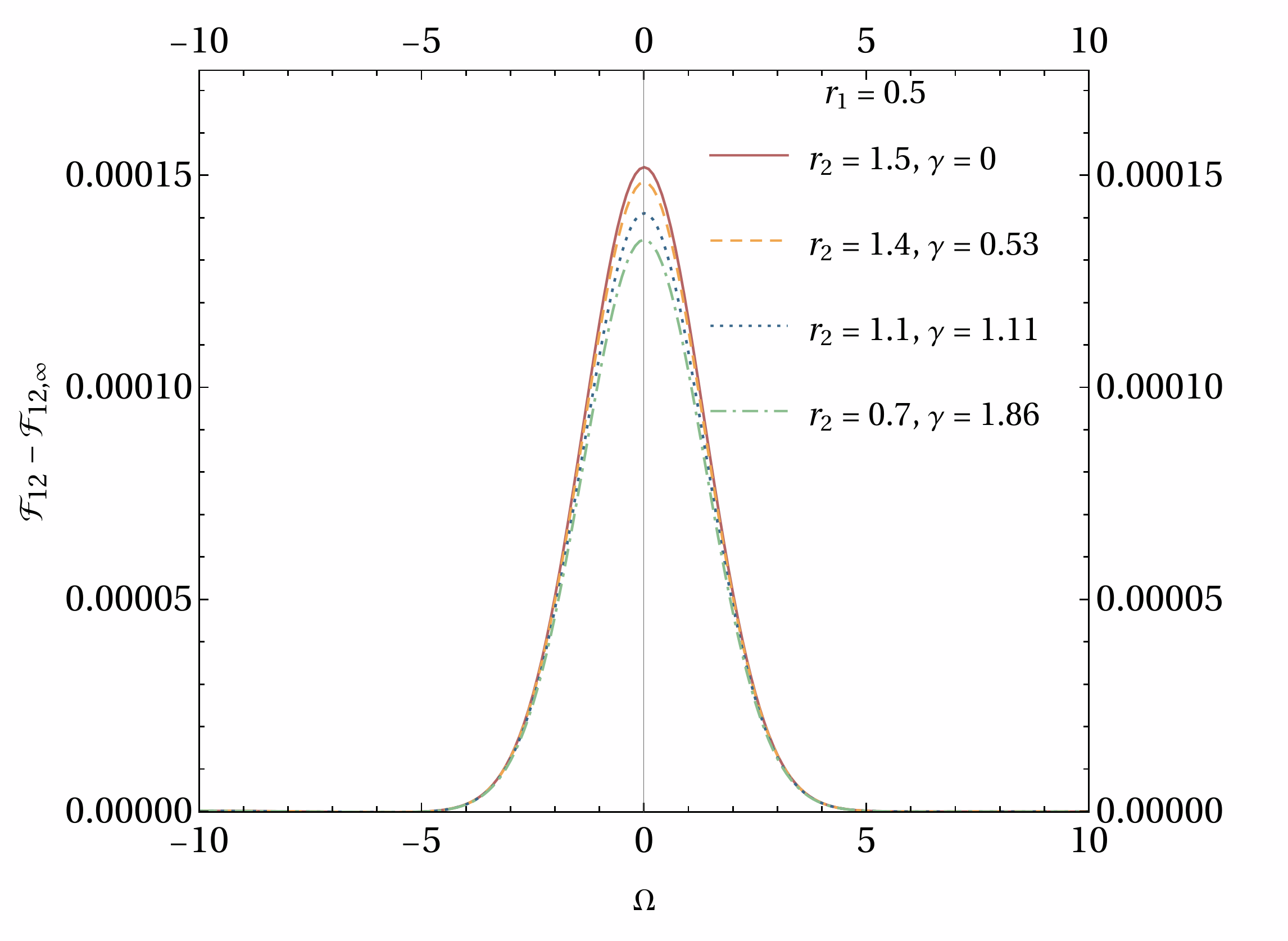}{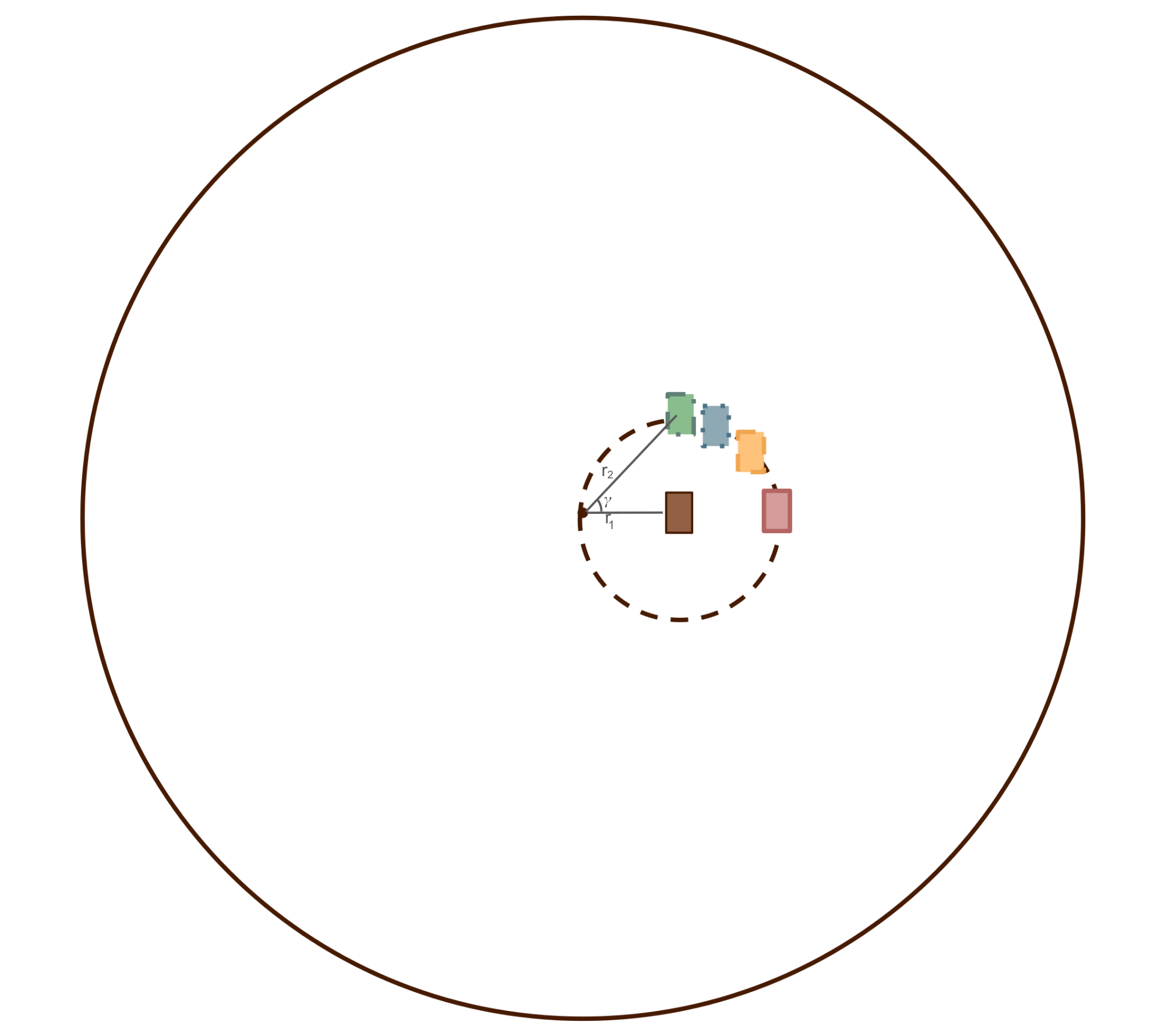}{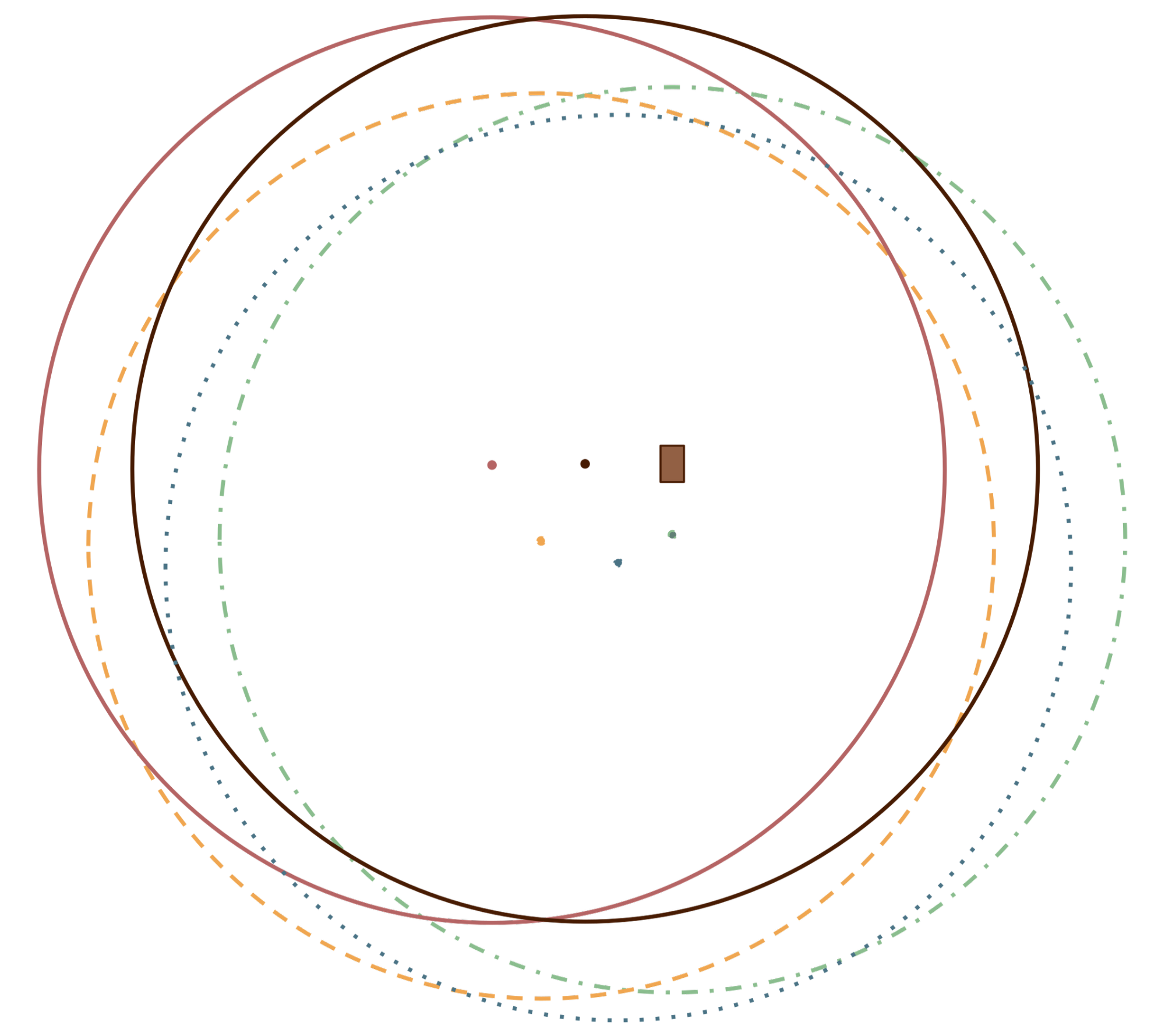}{Plots corresponding to $r_1 = 0.5$ and $r_2$ moving along a circle centered at $r_1$, with a fixed separation $|r_2 - r_1| = 1$. Panels (d1) and (d2) represent the frame with a fixed shell and the frame with the mass in superposition, respectively. Plot (a) shows the total response function, while panel (b) shows the cross-term $\mathcal{F}_{12}$. Plot (c) displays the correction relative to the limit where $R\rightarrow \infty$. As the distance between $r_1$ and $r_2$ remains constant, this setup allows for an analysis of how the proximity between the detector and the shell modifies the response function. }{fig:Conf3}{0 0 0 0}{1\textwidth}

The cross responses shown in Fig.~\ref{fig:Conf3}(b) almost coincide for these configurations. This indicates that, for the parameters considered here, the leading contribution to $\mathcal{F}_{12}$ is determined by the separation $D$ between the detector positions in the two branches of the superposition, rather than by the orientation of their relative separation or their position with respect to the center of the shell. This is expected from the flat-space limit, where translational and rotational invariance imply that the two-point function depends only on $D$. Since this distance is the same for all curves in Fig.~\ref{fig:Conf3}(b), the leading flat-space contribution is common to all configurations and dominates $\mathcal{F}_{12}$.

Subtracting the flat-space contribution $\mathcal{F}_{12,\infty}$ from $\mathcal{F}_{12}$ makes the difference between the response function in these configurations explicit, as shown in Fig.~\ref{fig:Conf3}(c). The magnitude of this difference is of order $10^{-4}$ and it represents the shell-induced corrections. The presence of the shell breaks translational invariance and modifies the field fluctuations in both branches, leading to corrections to the detector response. The cross-correlation and the total response function then depend not only on $D$ but are also sensitive to the positions of the detectors inside the shell. The effects induced by the shell become slightly larger as the detector position in the second branch moves farther from the center and, consequently, closer to the shell. Thus, this setup shows that the detector response depends not only on the separation between the two detector positions but also on the orientation and location of this separation within the shell.

From Fig.~\ref{fig:Conf3}(a), the same conclusion can be drawn for the total response $\mathcal{F}_{+}$, which is almost indistinguishable for the different orientations, showing that it is dominated by the flat-space contribution.

\begin{figure}
    \centering
    \includegraphics[width=\linewidth]{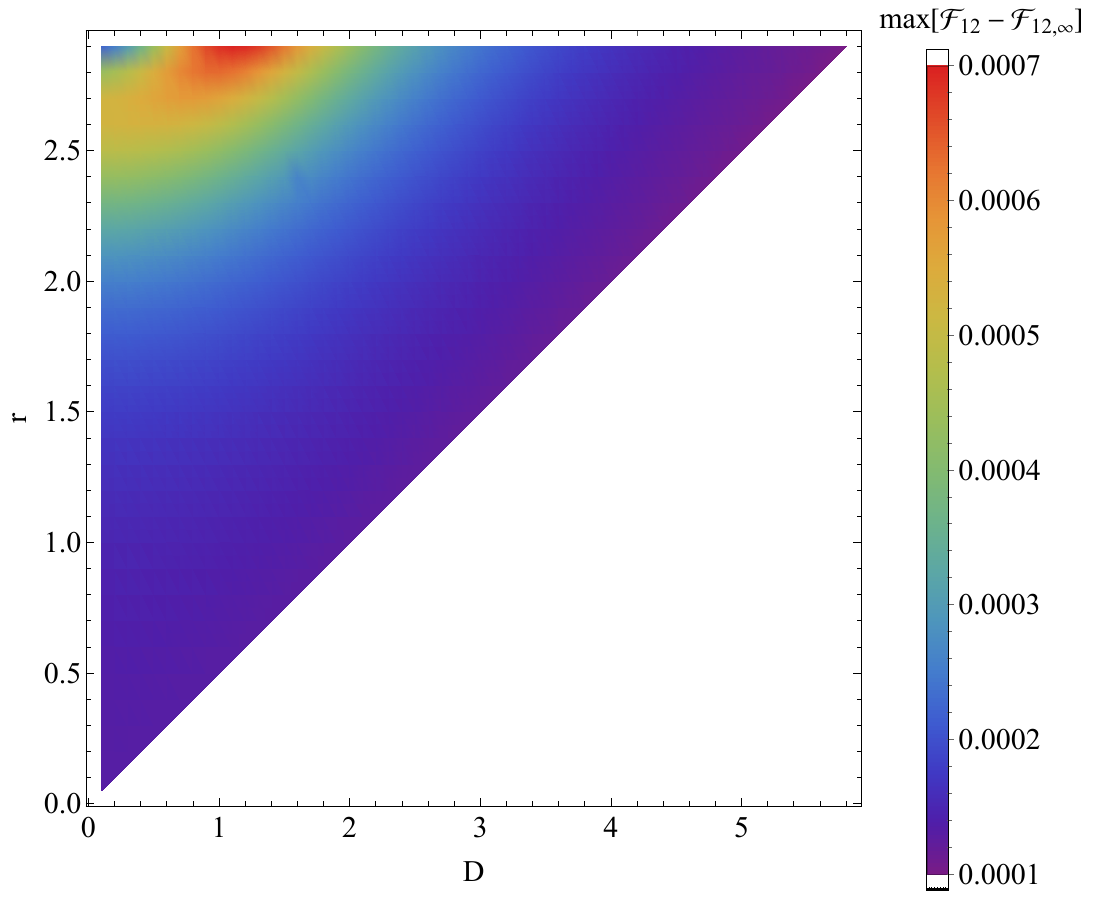}
    \caption{Maximum shell-induced correction to the cross-branch response, $\max[\mathcal{F}_{12}-\mathcal{F}_{12,\infty}]$, as a function of the distance $D$ between the detector positions and the common radial position $r=r_1=r_2$.  
}
    \label{fig:optimised_F12-F12flat}
\end{figure}

To identify configurations in which the interference between the branches due to the presence of the shell is more pronounced, we now compute the maximum of the difference $\mathcal{F}_{12}-\mathcal{F}_{12,\infty}$ over the energy gap $\Omega$ for a class of configurations parametrized by two free parameters. The radial coordinates of the detector positions in the two branches are taken to be the same, $r_1=r_2=r$, while their directions from the center of the shell can differ. The two detector positions are then separated by a distance $D$, with an angle $\gamma=2\sin^{-1}(D/2r)$ between their radial directions.  Since $D$ is a distance between two points at radius $r$, it lies in the interval $0 \leq D \leq 2r$. The results are shown in Fig.~\ref{fig:optimised_F12-F12flat}. The maximum shell-induced correction becomes more pronounced for larger values of $r$, reaching a maximum of $7.1 \times 10^{-4}$ at $r=2.9$ and $D=1.1$. For radii close to this value, the correction initially increases as $D$ increases from zero, reaches a maximum near $D\approx1$, and then decreases for further increases in $D$. For smaller radii well inside the shell, the correction is comparatively small and decreases as $D$ increases.

\begin{figure}
    \centering
    \includegraphics[width=\linewidth]{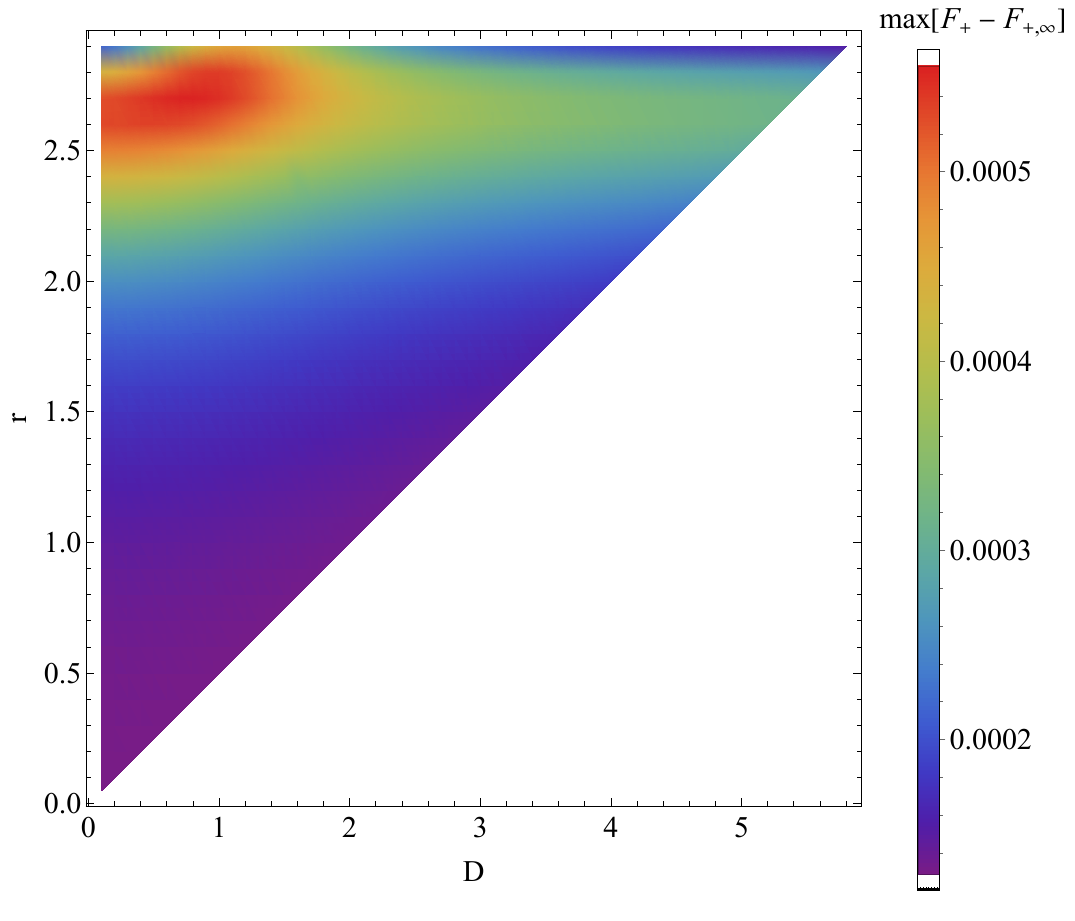}
    \caption{Maximum shell-induced correction to the total post selected response, $\max[F_{+}-F_{+,\infty}]$, as a function of the distance $D$ between the detector positions and the common radial position $r=r_1=r_2$. 
}
    \label{fig:optimised_F+-F+flat}
\end{figure}

We performed the same analysis for the maximum shell-induced correction to the total post-selected response ($\max[\mathcal{F}_{+}-\mathcal{F}_{+,\infty}]$). The results are shown in Fig.~\ref{fig:optimised_F+-F+flat}. The corrections exhibit similar general features but are more pronounced over a noticeably broader region. As in the previous case, configurations in which the detector positions are near the center are less affected by the presence of the shell, whereas $\max[\mathcal{F}_{+}-\mathcal{F}_{+,\infty}]$ increases significantly as both positions are moved closer to the shell. The maximum value of $\max[\mathcal{F}_{+}-\mathcal{F}_{+,\infty}]$ is $5.5 \times 10^{-4}$ at $r=2.7$ and $D=0.8$. In this case, the region of significant enhancement exists over a much wider range of values of $D$: even as the separation increases, a sizeable shell-induced contribution remains as long as the detector positions are sufficiently close to the shell. Hence, the total response is less sensitive to variations in the distance $D$ than the cross-branch correlation.

The post-selected response depends not only on the cross-branch correlation $\mathcal{F}_{12}$ but also on the same-branch correlations $\mathcal{F}_{ii}$. Our results indicate that, as $D$ increases, the shell-induced correlation between the branches becomes much weaker, while each branch can still exhibit significant shell-induced effects. This explains why the region of enhanced shell-induced response in Fig.~\ref{fig:optimised_F+-F+flat} extends over a wide range of separations $D$, whereas the corresponding region in Fig.~\ref{fig:optimised_F12-F12flat} is much more localized. Accordingly, we conclude that proximity to the shell alone is not sufficient to maximize the shell-induced cross-branch contribution: the separation between the two components of the spatial superposition must also be appropriately chosen.

\begin{figure}
    \includegraphics[width=.9\linewidth]{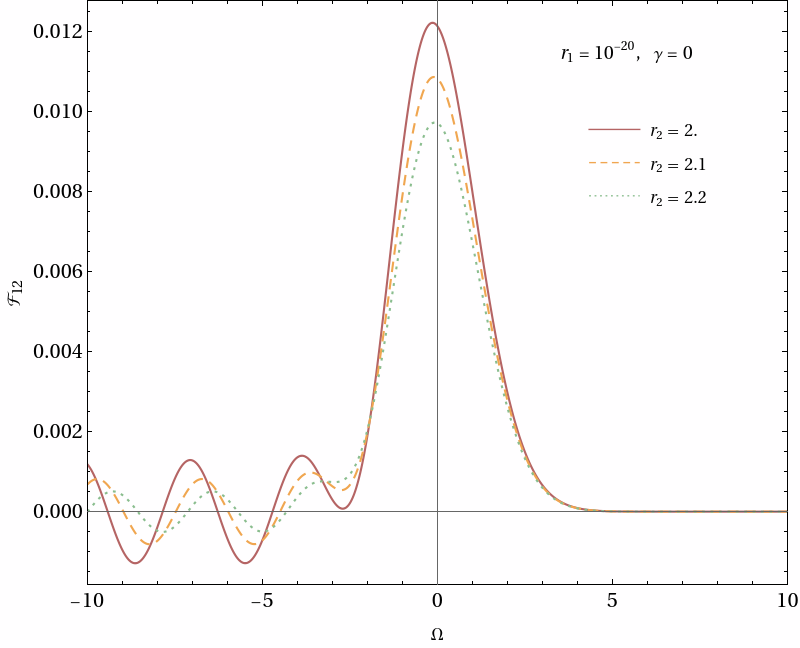}
    \caption{ Cross-branch response function $\mathcal{F}_{12}$ as a function of the energy gap $\Omega$ for $r_1 = 10^{-20}$, $\gamma = \pi$, and different values of $r_2$. As $r_2$ increases, the second local maximum becomes progressively less pronounced and, for certain values of $r_2$, ceases to define a distinct extremum of the response. This illustrates a change in the local peak structure of $\mathcal{F}_{12}(\Omega)$ as the separation between the detector positions increases.
}
    \label{fig:res_1}
\end{figure}

Finally, we analyze the relative height of the interference peaks observed in the cross-branch contribution $\mathcal{F}_{12}(\Omega)$. We find that the difference between the first two maxima, $i_1$ and $i_2$, becomes particularly pronounced for specific detector positions $r_1, r_2$.  This behavior is illustrated in Fig.~\ref{fig:res_1}, where the rapid decay of the interference peaks is clearly visible for a configuration with diametrically opposed positions with the same radial coordinate.

\begin{figure}
    \includegraphics[width=.8\linewidth]{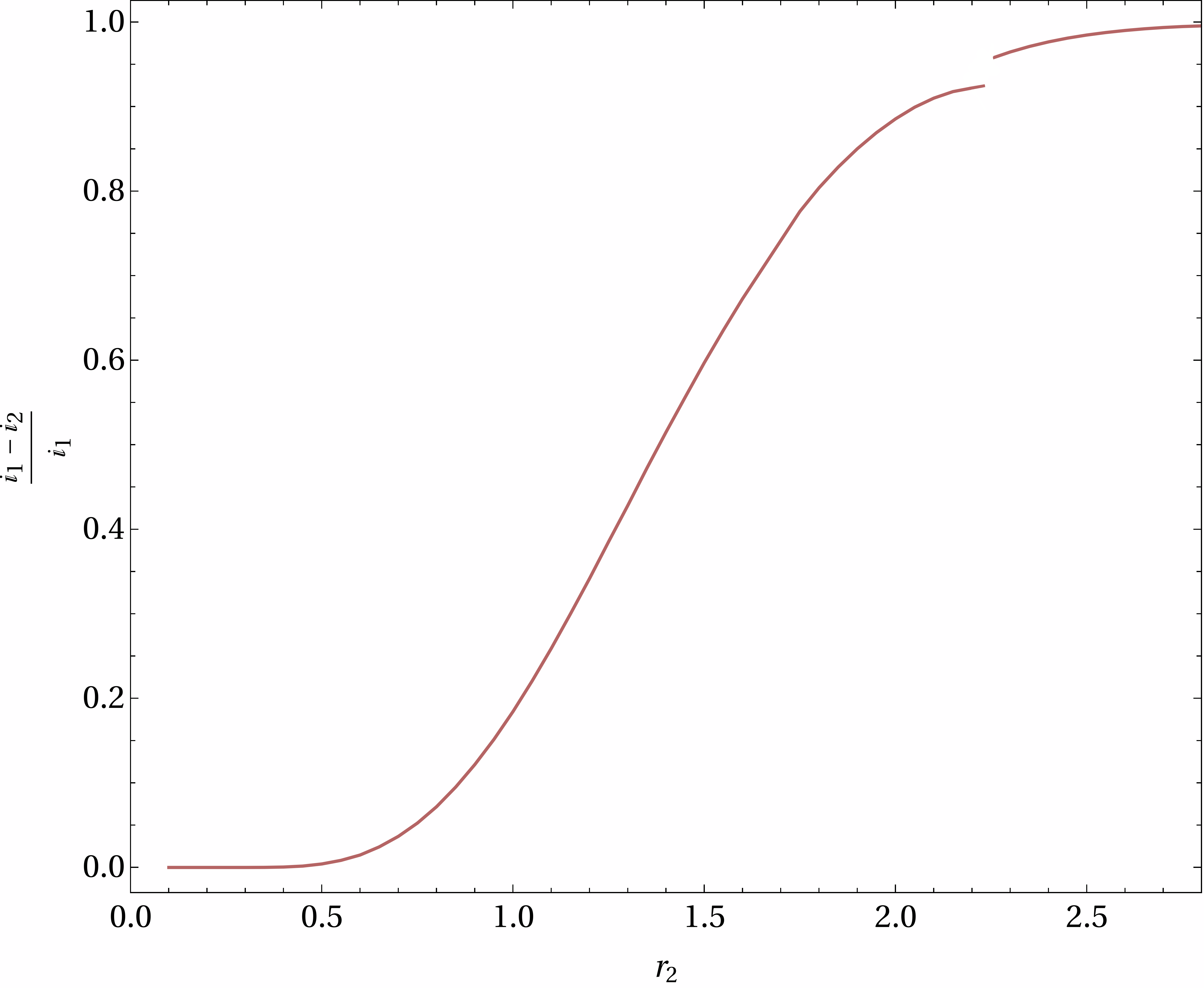}
    \caption{
Relative difference between the heights of the first two peaks of the cross-branch response function $\mathcal{F}_{12}$ as a function of $r_2$, for $r_1 = 10^{-20}$ and $\gamma = 0$. The peak heights $i_1$ and $i_2$ correspond, respectively, to the first and second peaks closest to $\Omega=0$. The sharp variation around $r_2 \simeq 2.2$ marks the disappearance of the second local maximum, as seen in Fig.~\ref{fig:res_1}, after which the subsequent maximum becomes the second peak in this ordering.
}
    \label{fig:res_2}
\end{figure}

To further analyze this effect, we plot in Fig.~\ref{fig:res_2} the normalized difference $(i_1-i_2)/i_1$ between the first two peaks closest to $\Omega=0$, $i_1$ and $i_2$, of the cross-branch correlation $\mathcal{F}_{12}(\Omega)$ for a class of configurations with $r_1=0$, $r_2$ variable and $\gamma=0$. This ratio increases as $r_2$ increases and the distance between the two positions grows. This behavior reflects a nonuniform suppression of the overall response: the intensity of the second peak, $i_2$, decreases substantially faster than that of the first peak, $i_1$, indicating an increasing dominance of the low-energy gap sector of the detector response. Around $r_2\simeq 2.2$, however, the local maximum associated with $i_2$ is progressively suppressed until it is no longer distinguishable as an independent peak. Beyond this point, the subsequent local maximum becomes the second peak closest to $\Omega=0$. The sharp variation observed in Fig.~\ref{fig:res_2} therefore reflects this change in the peak structure of $\mathcal{F}_{12}(\Omega)$, rather than a discontinuity in the response function itself.

\section{Discussion and conclusions}

A massive thin shell curves the surrounding spacetime, affecting the vacuum fluctuations of quantum fields.  These fluctuations can be probed by Unruh-DeWitt detectors, which can distinguish the vacuum state on a flat spacetime from that in the presence of the shell \cite{Ng:2016hzn}. We extend the analysis of UDW detectors in the gravitational field of a spherical mass shell in two directions: (i) we consider a detector in a quantum superposition of two trajectories, and (ii) we construct and analyze a model describing the interaction of a detector with the gravitational field of a spherical mass shell in a superposition of two positions. The second model is obtained by applying a quantum-reference-frame transformation to the first setup, assuming the invariance of physical observables, such as the detector excitation probability, under changes of quantum reference frame.

We numerically compute the response of a UDW detector located inside the shell using a Gaussian switching function. In the reference frame in which the mass is delocalized, the detector is located in a locally flat region common to both branches of the superposition. In the reference frame in which the mass is localized, this configuration corresponds to a detector in a superposition of two positions inside the shell. We obtain the following main results, described in the reference frame in which the detector is in a superposition of positions.

The superposition of positions introduces a new interference contribution $\mathcal{F}_{12}$ to the detector response that is absent for a detector at a definite position. As a result, an interference pattern is imprinted on the total post-selected response in the form of oscillations about the response corresponding to a classical mixture of the two branches of the superposition. For a fixed distance $D$ between the two possible detector positions, the excitation probability depends only weakly on the specific locations of the detector positions. Although the presence of the shell modifies the vacuum fluctuations, the dominant contribution to the detector response coincides with that in flat spacetime. This is consistent with the regime of weak gravitational fields considered here, in which the gravitational field produces only small corrections to the flat spacetime vacuum fluctuations.

To isolate the effects of the shell, we analyze the difference $\mathcal{F}_{12}-\mathcal{F}_{12,\infty}$ between the interference contribution to the detector response in the presence of the shell and its flat-spacetime counterpart. We find that this difference is maximized at vanishing detector energy gap, $\Omega=0$, and that its maximum value is generally three orders of magnitude smaller than that of the interference contribution itself. We present an extensive analysis of how the shell-induced corrections to the interference term $\mathcal{F}_{12}-\mathcal{F}_{12,\infty}$ and to the total response $\mathcal{F}_+ -\mathcal{F}_{+,\infty}$ depend on the detector positions, and identify configurations for which these corrections are maximized.

In the reference frame in which the mass is delocalized, the coherent superposition of geometries is reflected in the interference contribution $\mathcal{F}_{12}(\Omega)$, which gives rise to oscillations in the post-selected detector response $\mathcal{F}_+(\Omega)$. These oscillations constitute the main signature distinguishing the detector response in the presence of a mass shell in a coherent superposition of positions from that of a shell at a definite position. In general, the curve of $\mathcal{F}_{12}(\Omega)$ exhibits oscillations for negative values of $\Omega$, reaching its highest peak near $\Omega=0$, before decreasing toward negligible values for positive $\Omega$. We identify particular configurations in which the highest peak is substantially more pronounced than the others. We find that, within the parameter range explored, the shell-induced corrections to the interference term, $\mathcal{F}_{12}-\mathcal{F}_{12,\infty}$, are maximized for separations between the shell positions of order $D \sim R/3$, where $R$ is the radius of the shell. These corrections decrease for larger separations, in contrast to other effects involving superpositions of geometries which are enhanced as the separation between the mass positions increases. 

The use of quantum-reference-frame transformations to describe quantum systems in nonclassical geometries generated by masses in superpositions of positions has been explored in previous works \cite{ACVik2023, CepollaroBTZ2026, Castro-Ruiz:2019vsu}. This provides a concrete method for investigating quantum gravity effects in setups in which a change of quantum reference frame can be implemented, as is the case of a superposition of mass distributions related by spatial translations. To our knowledge, the present work is the first application of this approach for a quantum field theory in four spacetime dimensions in the gravitational field produced by a source mass in a superposition of positions.

In contrast to quantum field theory on a definite classical spacetime, there is no standard procedure for defining field operators or the Hilbert space of a quantum field in a superposition of geometries. In particular, the description of interference terms arising from a coherent superposition of geometries goes beyond the framework of quantum field theory in curved spacetimes. Under a change of quantum reference frame, the contributions of these terms to the response function of a UDW detector are here mapped to corresponding interference terms for a detector in a superposition of positions on a definite geometry. In our approach, this correspondence provides a prescription for defining the matrix elements of products of field operators that enter the interference terms associated with a superposition of geometries.

Natural extensions of this work could explore other configurations involving superpositions of mass distributions, such as massive shells with distinct radii, or more general superpositions of nonisometric spacetimes. It is unclear whether an approach based on a change of quantum reference frame could be employed in these cases, and a distinct approach may be required. A more conceptual question related to our work concerns the rigorous construction of field operators and associated quantum states in the superpositions of geometries commonly considered in quantum gravity phenomenology. In particular, it would be interesting to investigate whether changes of quantum reference frame could be applied within the formalism of algebraic quantum field theory to construct quantum field models for simple superpositions of geometries.

In the final stages of this paper, we became aware of independent work by S. Gauvin and R. B. Mann, titled “Coherent detector responses to quantum-controlled spherical-shell geometries”. They investigate an Unruh-DeWitt detector inside a massive shell in a superposition of locations, of radii, or both. Using a common incoming-mode identification of the branch field algebras, their analysis examines interference in the detector response and the role of branch-dependent detector-clock phases. That work was submitted to arXiv on the same date as the current manuscript.

\section{Acknowledgement}

NSM thanks Joshua Foo for insightful discussions. SSS and NSM thank Keith Ng and Robert Mann for helpful discussions about the codes of reference~\cite{Ng:2016hzn}. SSS acknowledges funding from the Doktogrant Shariff S - 22019905. SSS and NSM acknowledge funding from QUAS VEGA 2/0164/25. NSM acknowledges funding from the project INDORBLE 09I03-03-V04-00679 for excellent researchers R2.
This work was carried out with the support of the Italian Ministry of University and Research (MUR) under the Funding for Research Projects (FIS 2), pursuant to Ministerial Decree No. 23314 of 11/12/2024 (Project Q-GraSp, FIS-2023-02629). This work was made possible through the support of the WOST, WithOut SpaceTime project (https://withoutspacetime.org), led by the Center for Spacetime and the Quantum (CSTQ), and supported by Grant ID\#~63683 from the John Templeton Foundation (JTF). The opinions expressed in this work are those of the author(s) and do not necessarily reflect the views of the John Templeton Foundation or those of the Center for Spacetime and the Quantum.
The authors benefited from the activities of COST Actions CA23115: Relativistic Quantum Information (https://rqi-cost.org/) and CA23130: Bridging high and low energies in search of quantum gravity (https://web.infn.it/BridgeQG/), funded by COST (European Cooperation in Science and Technology).

\section{Appendix}
\subsection{Radial modes}\label{appendix:radial_modes}
Let us derive the radial modes for a single classical massive shell. The field modes $\Psi_{\omega lm}$ satisfy the Klein-Gordon equation given in Eq.~\eqref{eq:KG_def}. For a metric of the form
\begin{equation}
ds^2 = - \alpha(r)^2 dt^2 + a^2(r) dr^2 + r^2 d\Omega^2 \,, 
\end{equation}
it can be explicitly written in terms of the metric components as \begin{multline}\label{eq:KG_spherical_form}
    -\frac{1}{\alpha^2} \partial_t^2 \Psi+\frac{1}{\alpha a r^2} \partial_r\left(\frac{\alpha}{a} r^2 \partial_r \Psi\right) \\ 
    +\frac{1}{r^2 \sin \theta} \partial_\theta\left(\sin \theta \partial_\theta \Psi\right)+\frac{1}{r^2 \sin ^2 \theta} \partial_\phi^2 \Psi = 0 \, .
\end{multline}

The metric is spherically symmetric, which implies that the modes factorize as the product of a radial and an angular function. The angular part can be expanded in spherical harmonics. Accordingly, we write the normal modes as 
\begin{equation}
    \Psi_{\omega l m} = \frac{1}{\sqrt{4\pi\omega}}e^{-i\omega t} \psi_{\omega l}(r) Y_{lm}(\theta, \phi).
\end{equation}
The spherical harmonics are eigenfunctions of the angular part of Eq.~\eqref{eq:KG_spherical_form},
\begin{multline}
    \Big[ -\frac{1}{\sin \theta} \partial_\theta \Big( \sin \theta \partial_\theta \Big) - \frac{1}{ \sin^2 \theta} \partial_\phi^2 \Big] Y_{lm}(\theta, \phi) \\
    = l(l+1)\ Y_{lm}(\theta, \phi),
\end{multline}
with $l=0,1,2,3,...$ Using these results, we can simplify Eq.~\eqref{eq:KG_spherical_form} and rewrite it in a form that helps us focus on the behavior of the radial modes, which are modified due to the presence of a shell. We find that the radial mode satisfies:

\begin{equation}
    \omega^2 \psi_{\omega l}+\frac{\alpha}{a r^2} \partial_r\left(\frac{\alpha}{a} r^2 \partial_r \psi_{\omega l}\right)-\alpha^2 \frac{l(l+1)}{r^2} \psi_{\omega l}=0 \, .\label{eq:KG_eqn_radial_modes_psi}
\end{equation}

We can recast the above equation into a more familiar form by introducing the tortoise coordinate $r^*$ such that
\begin{equation}
    \partial_{r^*}=\frac{\alpha}{a} \partial_r
\end{equation}
and defining
\begin{equation}
\rho_{\omega l} \equiv r \psi_{\omega l} .
\end{equation}
This yields:

\begin{equation}
    \partial_{r^*}^2 \rho + \Big[ \omega^2 - {V_l} (r) \Big] \rho =0  \, ,
    \label{eq:KG_scattering_form}
\end{equation}
where the effective potential $V_l(r)$ is given by 
\begin{align}
    V_l (r) =& \frac{1}{r} \partial_{r^*}^2 r + \alpha^2 \frac{l(l+1)}{r^2}\notag\\
            =& \frac{1}{r} \Big( \frac{\alpha}{a} \Big) \partial_{r} \Big( \frac{\alpha}{a} \Big) + \alpha^2 \frac{l(l+1)}{r^2}.\label{eq:effective_potential} \quad\because \partial_{r^*}r=\frac{\alpha}{a}
\end{align}
Eq.~\eqref{eq:KG_scattering_form} has the characteristic form of a one-dimensional scattering problem in quantum mechanics. Using the expressions of $\alpha$ and $a$ from the metric defined in Eq.~\eqref{eq:metric_def} and substituting them in the effective potential defined in Eq.~\eqref{eq:effective_potential} we get
\begin{equation}
V_l(r)= \begin{cases}f(r)\left[\frac{l(l+1)}{r^2}+\frac{2 M}{r^3}\right], & r>R \label{eq:effective_potential_r>R}\\ f(R)\left[\frac{l(l+1)}{r^2}\right], & r<R .\end{cases}
\end{equation}

In the region inside the shell, the Eq.~\eqref{eq:KG_scattering_form} simplifies and assumes the form:

\begin{equation}
    \tilde{r}^2 \psi + 2 \tilde{r} \partial_{\tilde{r}} \psi + \tilde{r}^2 \partial_{\tilde{r}}^2 \psi - l(l+1) \psi = 0 \, ,
    \label{eq:spherical_bessel_fn}
\end{equation}
with $\tilde{r}=\omega r^*$ and $\partial_{\tilde{r}}=\omega^{-1} \partial_{r^*}$. This is nothing but the spherical Bessel differential equation. Therefore, the solutions to the radial part of the Klein-Gordon equation within the shell, where the metric is isometric to a patch of Minkowski spacetime, are given by $\psi \sim j_l\left(\omega r^*\right)=j_l(\tilde{\omega} r)$ where $\tilde{\omega} \equiv \omega / \sqrt{f(R)}$. In the region exterior to the shell, where the metric is isometric to a patch of Schwarzschild spacetime, solutions are obtained by numerically solving the differential equation\ (\ref{eq:KG_scattering_form}) with the effective potential given in Eq.~\eqref{eq:effective_potential_r>R} for $r>R$. The interior and exterior solutions are then glued at the shell ($r=R$) by matching conditions: the mode function must be continuous at the boundary, $\psi\big|_{r=R}=j_l(\tilde{\omega} R)$, and its derivative with respect to the tortoise coordinate must also match, $\partial_{r^*} \psi\big|_{r=R}=\frac{\alpha}{a} j_l^{\prime}(\tilde{\omega} R)$. In other words, we use the values of $\psi\big|_{r=R}$ and $\partial_{r^*} \psi\big|_{r=R}$ as initial conditions for solving the outer differential equation.

The normalization condition for the modes reads (see the appendix of \cite{Ng:2016hzn} for the derivation):
\begin{equation}\label{eq:modes_normalization_cond}
\int_0^{\infty} \mathrm{d} r^* \rho_{\omega_1 l_1}(r) \rho_{\omega_2 l_2}(r)=2 \pi \delta\left(\omega_1-\omega_2\right) \delta_{l_1, l_2} \, .
\end{equation}  
The inner product is dominated by the behavior of the modes for large radii, allowing the region inside the shell to be neglected in the integration. Under this approximation, the normalization condition implies an asymptotic behavior for the modes of the form: $\rho_{\omega l}(r) \rightarrow 2 \sin (\omega r+\theta)$ as $r \rightarrow \infty$. As a result, if $\tilde{\psi}_\text{out}$ is the numerically determined exterior solution, then the normalization condition requires that $A_{\omega l}\tilde{\psi}_\text{out}\rightarrow 2 \sin(\omega r)/r$ as $r \rightarrow \infty$. This requirement fixes the normalization constant $A_{\omega l}$, completing the determination of the normal modes.

\subsection{Extra configuration}\label{extraconfig}

\manyfiguresb{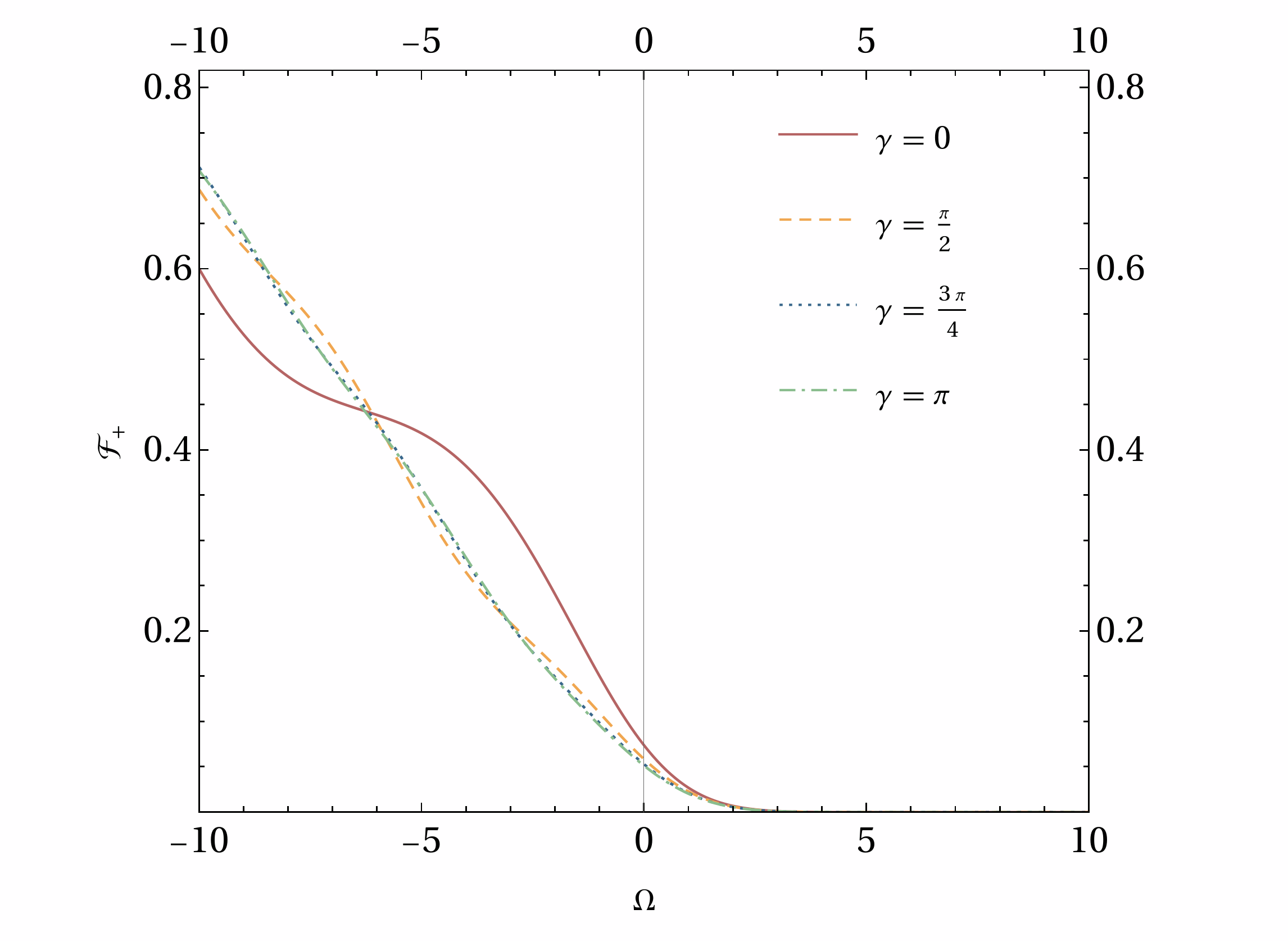}{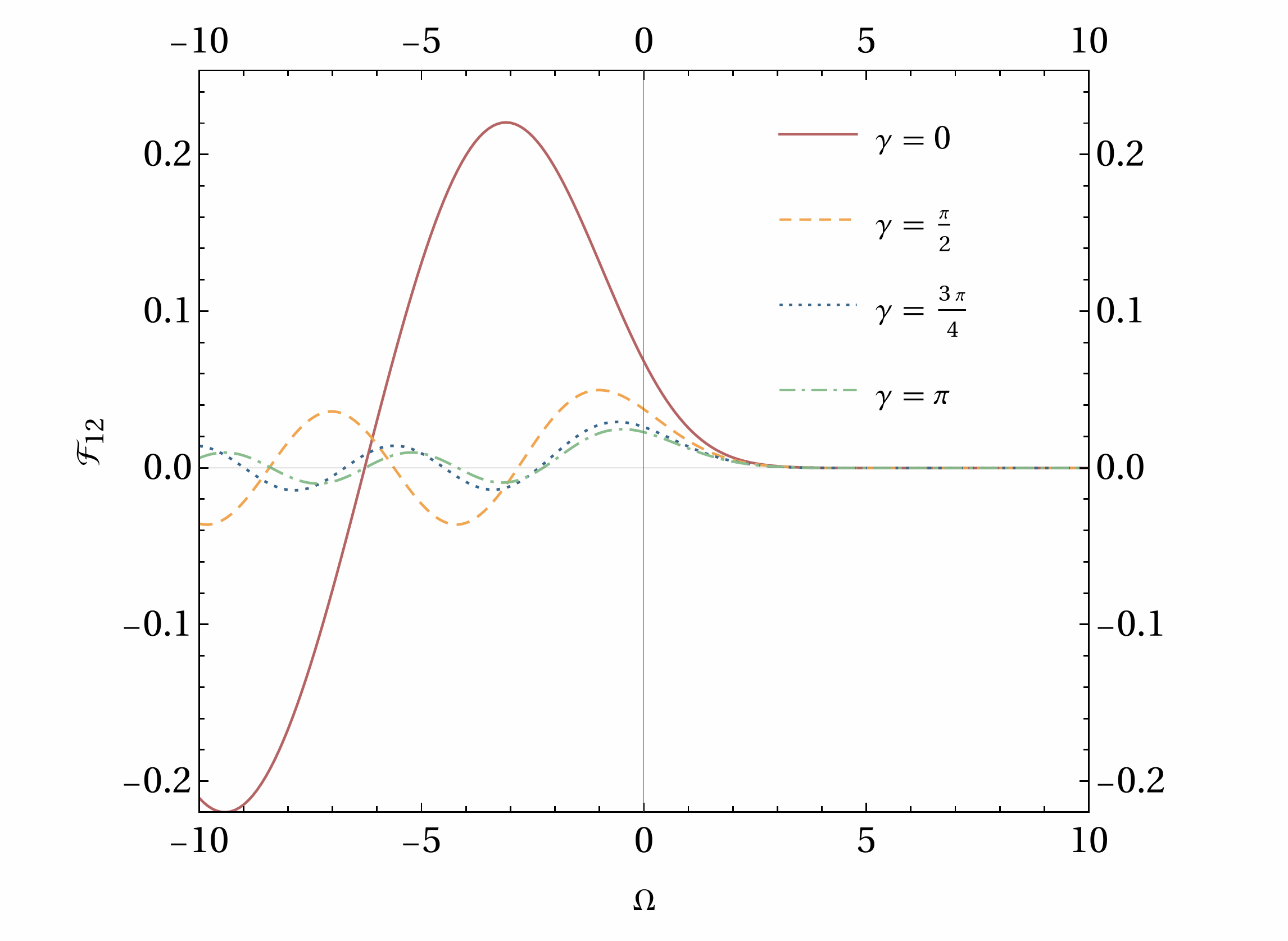}{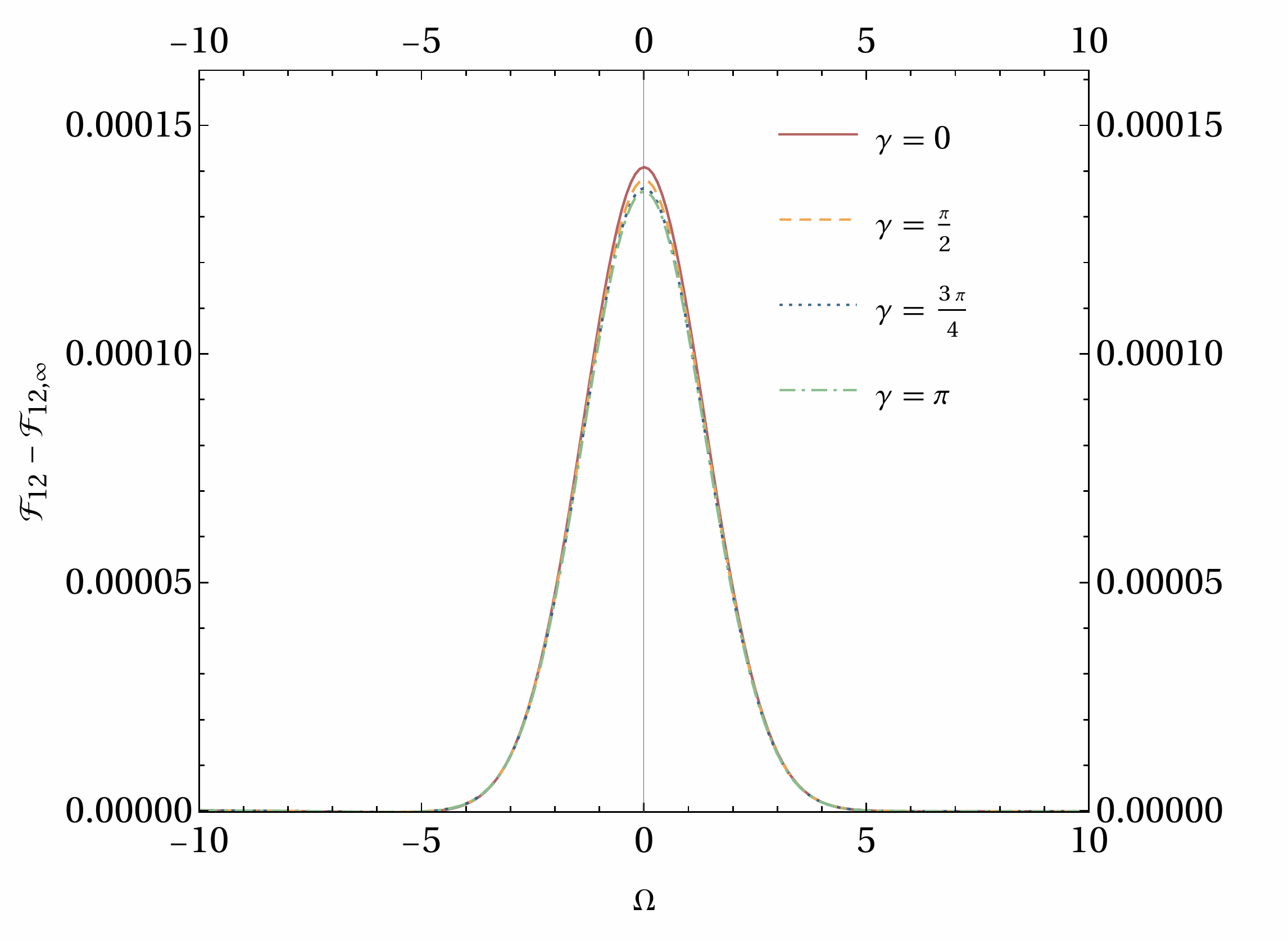}{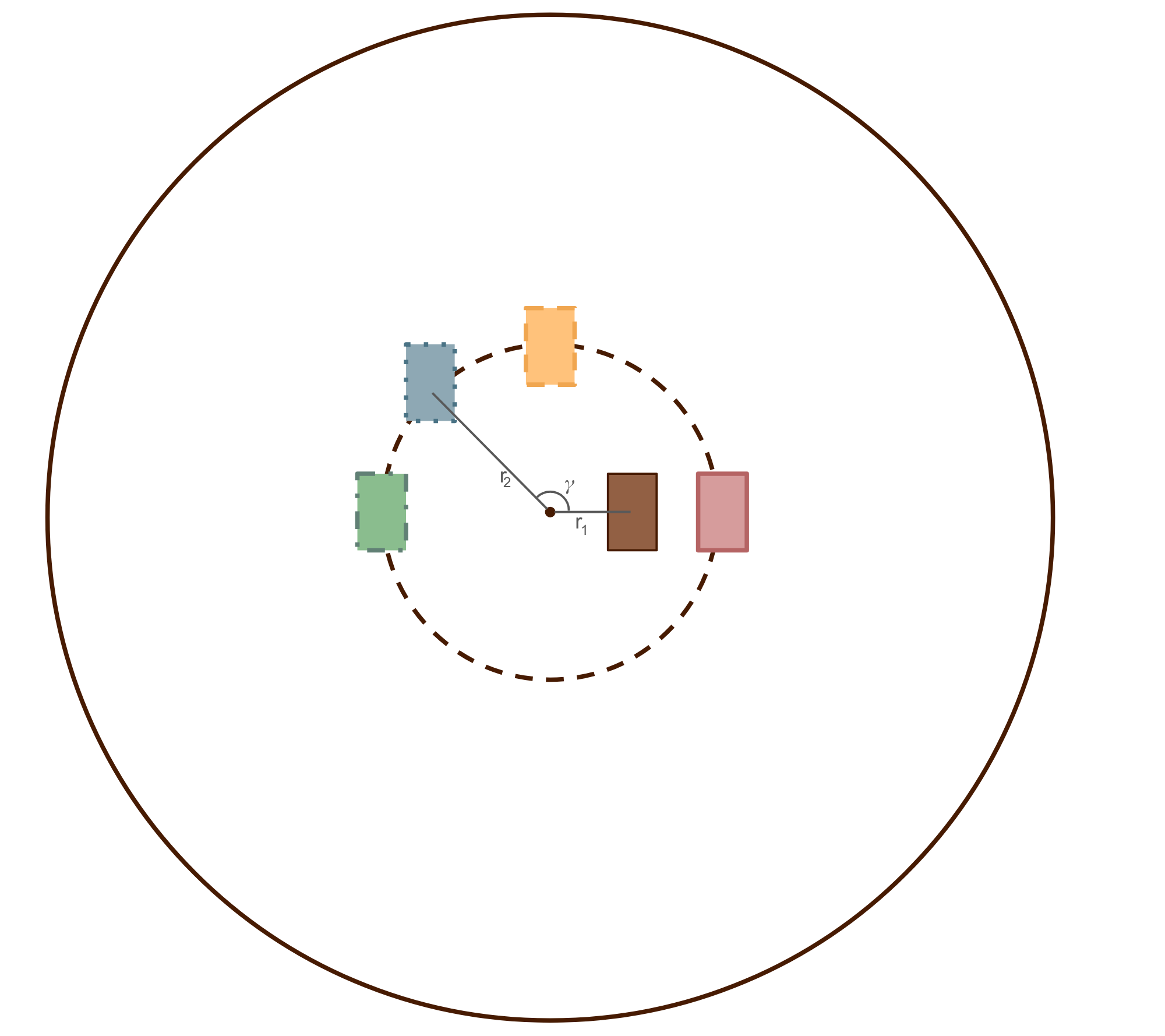}{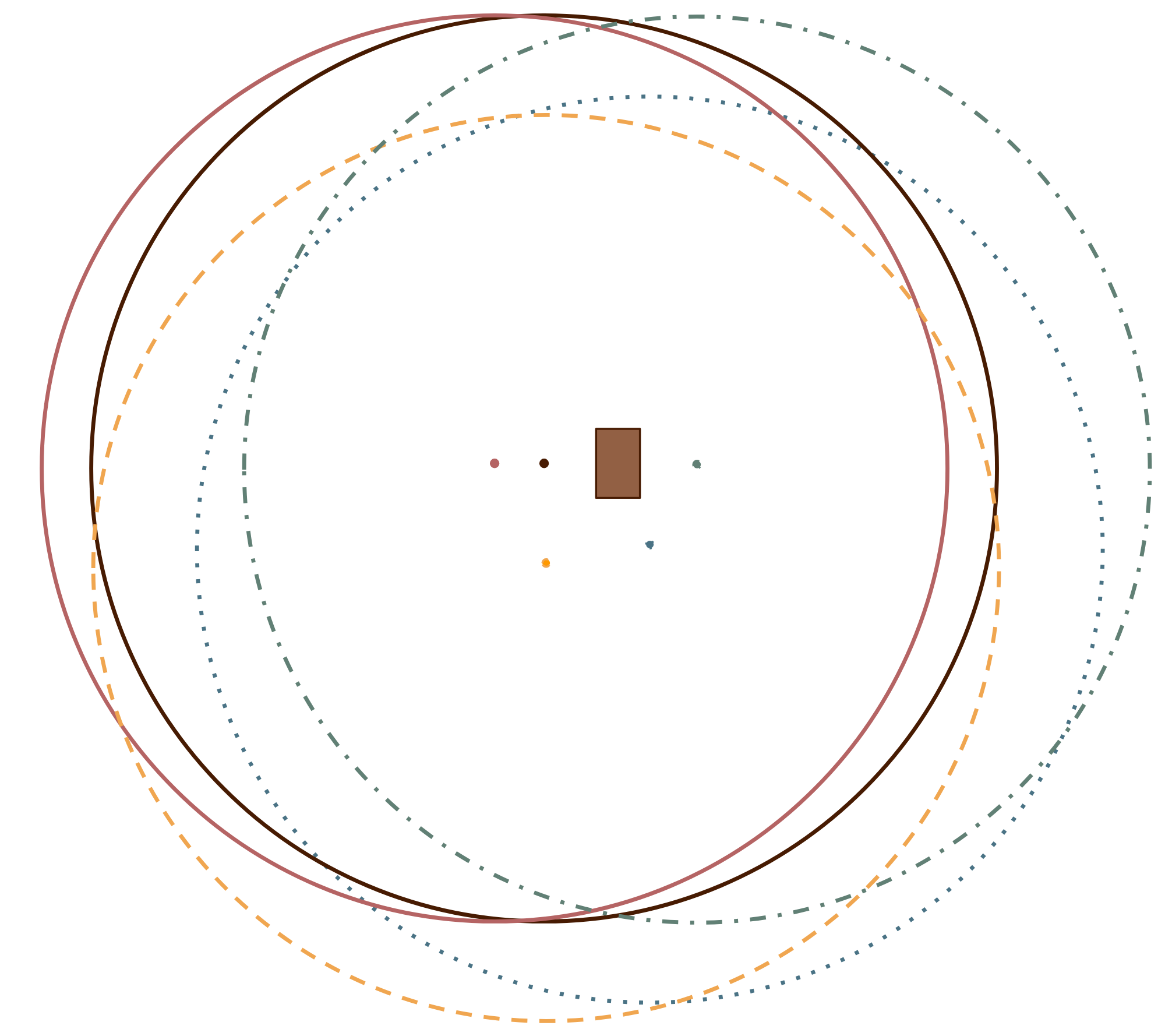}{Plots corresponding to $r_1 = 0.5$ and $r_2$ moving along a circle of radius $1.0$ concentric to the shell. Panels (d1) and (d2) correspond to the frame where the shell is at a fixed location and the frame where the mass is in a superposition of positions, respectively. Plot (a) shows the total response function, while panel (b) corresponds to the cross-term $\mathcal{F}_{12}$. Plot (c) shows the difference between the shell of radius $R=3$ and the infinitely large shell limit ($R\rightarrow \infty$). These configurations maintain a fixed distance between the detector branches and the shell, thus providing an analysis of how the separation between the two detector positions interferes with the response function. }{fig:Conf2}{0 0 0 0}{1\textwidth}

In this section, we describe results for an additional class of configurations for the detector positions in the reference frame in which the massive shell is localized. In the first branch, we consider a fixed position with $r_1=0.5$, while in the second branch we choose $r_2=1$ and vary the angular coordinate, covering a circle of unit radius centered at the origin. Since each position remains at a fixed radial coordinate, the distance of each position to the shell is fixed for these configurations, while the separation $D$ and the angle $\gamma$ between the detector positions are varied. These quantities are related by:
\begin{equation}
  D=|\mathbf{r}_2 - \mathbf{r}_1|
  = \sqrt{r_1^2 + r_2^2 - 2 r_1 r_2 \cos\gamma}.
\end{equation}

Since the same-branch contributions $\mathcal{F}_{11}$ and $\mathcal{F}_{22}$ depend only on the fixed radial positions, all of the $\gamma$ dependence of the total post-selected response $\mathcal{F}_{+}$, shown in Fig.~\ref{fig:Conf2}(a), comes from the cross-branch term $\mathcal{F}_{12}$, shown in Fig.~\ref{fig:Conf2}(b). The $\gamma = 0$ curve, with the smallest separation $D$ between the positions among these configurations, displays the largest peak amplitude in $\mathcal{F}_{12}$. Around the value of $\Omega$ where this peak occurs, the corresponding $\mathcal{F}_{+}$ curve has a broad shoulder, which reflects the coherent addition of $\mathcal{F}_{12}$ to the same-branch terms $\mathcal{F}_{11}$ and $\mathcal{F}_{22}$.

To isolate the effects due to the presence of the shell in the cross-branch correlations, we determine the difference $\mathcal{F}_{12} - \mathcal{F}_{12,\infty}$. The results are shown in Fig.~\ref{fig:Conf2}(c). In contrast to the cross-branch correlation $\mathcal{F}_{12}$, shown in Fig.~\ref{fig:Conf2}(b), this shell-induced correction depends only weakly on $\gamma$. The strong dependence of $\mathcal{F}_{12}$ on the separation is therefore contained almost entirely in the flat-space function $\mathcal{F}_{12,\infty}$ of reference, and varying $D$ at fixed radial positions changes the shell-induced correction only slightly, with the value of the peak decreasing marginally as $D$ increases.

These results, together with those represented in Fig.~\ref{fig:Conf3}, clarify the behavior seen in Fig.~\ref{fig:Conf1}(c). In the configurations represented in Fig.~\ref{fig:Conf1}, the separation between the detector positions and the proximity of the position in one of the branches to the shell vary simultaneously, since $d = r_2$. The growth of the correction with $r_2$ cannot therefore be attributed to the variation of either of these properties independently, solely from the analysis of Fig.~\ref{fig:Conf1}(c). The setups of Figs.~\ref{fig:Conf2} and~\ref{fig:Conf3} allow us to distinguish between the two effects. When the radial positions are held fixed and only $D$ changes (Fig.~\ref{fig:Conf2}(c)), the peak of the correction displays only a small variation. When, instead, the separation is held fixed at $d = 1$, and one of the positions is moved closer to the shell (Fig.~\ref{fig:Conf3}(c)), the correction grows with $r_2$ and has a noticeably larger variation. These comparisons suggest that the increase seen in Fig.~\ref{fig:Conf1}(c) is predominantly associated with one of the positions approaching the shell. The comparison is qualitative, since the inner position lies at the centre of the shell in Fig.~\ref{fig:Conf1} but at $r_1 = 0.5$ in Figs.~\ref{fig:Conf2} and~\ref{fig:Conf3}. Moreover, these comparisons concern positions located well inside the shell. As shown in Fig.~\ref{fig:optimised_F12-F12flat}, for $r_1 = r_2 = r$, the correction becomes strongly sensitive to $D$ when the branches approach the boundary.

\setlength{\bibsep}{4pt}
\bibliographystyle{apsrev4-2}
\bibliography{bibliography}

\end{document}